\documentclass[12pt]{article}
\usepackage{epsf}
\def\G{{\cal G}^{+++}}
\begin{document}
\thispagestyle{empty}
\setcounter{page}{0}
\renewcommand{\theequation}{\thesection.\arabic{equation}}

{\hfill{ULB-TH/03-14}}

{\hfill{DCPT-03/21}}

{\hfill{KCL-MTH-03-04}}

{\hfill{\tt hep-th/0304206}}

\vspace{.5cm}

\begin{center} {\bf THE SYMMETRY OF M-THEORIES}

\vspace{.5cm}

Fran\c cois Englert${}^a$, Laurent Houart${}^b$\footnote{Research  
Associate
F.N.R.S.}, Anne Taormina${}^c$
\\ and Peter West${}^d$

\footnotesize \vspace{.5 cm}

${}^a${\em Service de Physique Th\'eorique\\ Universit\'e Libre de  
Bruxelles,
  Campus Plaine, C.P.225\\Boulevard du Triomphe, B-1050 Bruxelles,  
Belgium}\\
fenglert@ulb.ac.be

\vspace{.2cm}

${}^b${\em Service de Physique Th\'eorique et Math\'ematique }\\  {\em
Universit\'e Libre de Bruxelles, Campus Plaine C.P. 231}\\ {\em  
Boulevard du
Triomphe, B-1050 Bruxelles, Belgium}\\ lhouart@ulb.ac.be

  ${}^c${\em Department of Mathematical Sciences}, {\em University of  
Durham}\\
{\em South Road, DH1 3LE Durham, U.K.}\\anne.taormina@durham.ac.uk

\vspace{.2cm}
  ${}^d${\em Department of Mathematics}, {\em King's College}\\{\em  
London,
U.K.}\\pwest@mth.kcl.ac.uk
\end{center}
\vspace{.2cm}
\begin{center}
\small {\bf Abstract}
\end{center}
\begin{quote}
\small
  We  consider the Cartan subalgebra of any very extended
  algebra ${\cal G}^{+++}$ where ${\cal G}$ is a simple Lie algebra and  
let the
parameters be  space-time fields. These are identified with diagonal  
metrics and
dilatons. Using the properties of the algebra, we find that for all  
very
extensions
${\cal G}^{+++}$ of simple Lie algebras  there are theories of gravity  
and matter,
which admit classical solutions carrying representations of  the Weyl  
group  of
${\cal G}^{+++}$.  We also identify   the $T$ and $S$-dualities of  
superstrings and
of the bosonic string  with Weyl reflections and outer automorphisms of
well-chosen very extended algebras and we exhibit specific features of  
the
very extensions. We take these results as  indication that very  
extended
algebras underlie symmetries of any consistent theory of gravity and  
matter,
and might  encode  basic information for the construction of such  
theory.
\end{quote}

\setcounter{footnote}{0}
\newpage

\setcounter{equation}{0}
\section{Introduction}
\normalsize There is a widespread belief that supersymmetry and string   
theory
are essential ingredients in a single unified theory of physics.  
However,
  supergravity  theories do not provide a consistent theory of quantum  
gravity
and there exists no satisfactory theory of strings  in the sense that  
we can
calculate, even as a matter of principle, non-perturbative effects in  
non-trivial
backgrounds. Much of what we know about superstring theory outside
perturbation theory has been derived from the properties of the maximal
supergravity theories in ten dimensions, namely the IIA supergravity
\cite{campbellw84,huqn85,gianip84} and IIB supergravity
\cite{schwarzw83,howew84,schwarz83} theories as  well as the type I
supergravity coupled to the Yang-Mills theory \cite{brinkss77}.  These  
theories
are essentially determined by  the type of supersymmetry they possess  
and
hence they are the  complete low energy effective actions of the   
superstring
theories with the corresponding space-time supersymmetry.  Although the
existence of a unique supergravity theory in eleven dimensions
\cite{cremmerjs78} has been known for many years, it is only relatively  
recently
that it has been conjectured \cite{townsend95,witten95} to be a  
particular limit
of  a fully quantum theory called M-theory which would encompass all
superstring theories. Very little is known for sure about M-theory, but  
it   is
thought to contain, in particular limits and at low energies,   all  the
supergravity theories mentioned above.
\par
It is our feeling that essential  elements are lacking in the quest
for   a unified
theory of quantised gravity and matter.  Bringing to light the  
symmetries that
lie beneath the present attempts to  such unification might help  
uncover those
essential elements, which could actually hide in the symmetry structure  
itself. These hopes have been encouraged by some developments in
recent years that have argued that certain types of Kac-Moody algebras
occur in supergravity theories. Based on the formulation of eleven
dimensional supergravity as a non-linear realisation \cite{west00}, it
has been conjectured that an extension of this theory possesses a rank
eleven Kac-Moody algebra called $E_{11}$ \cite{west01}. Furthermore, it 
has been argued  that a rank twenty-seven Kac-Moody algebra, called
$K_{27}$ underlies the closed bosonic string in twenty-six dimensions 
\cite{west01} and that even pure gravity  in D
dimensions can be extended so as to possess a Kac-Moody algebra of rank
D \cite{lambertw01}. A systematic account of symmetries in
supergravity and string theories is given at the end of this section. 
\par
One interesting aspect of the proposed Kac-Moody symmetries for  
M-theory, for
the closed bosonic string and for gravity, is that they are all  
Kac-Moody algebras
of a special type. Given any finite-dimensional simple Lie algebra  
${\cal
G^{}}$,  there is a well-known procedure for   constructing  a  
corresponding
affine algebra ${\cal G^{+}}$ by adding a node  to the Dynkin diagram  
in a certain
way which is related to the properties of the highest root of ${\cal  
G^{}}$. One
may also further increase by one the rank of the algebra ${\cal G^{+}}$  
  by
adding  to the Dynkin diagram a further node that is attached to the   
affine node
by a single  line
\cite{goddardo84}. This is called the overextension  ${\cal G^{++}}$.  
The very
extension, also called triple extension in this paper, denoted ${\cal G^{+++}}$, is
found by adding yet another   node to the
Dynkin diagram that is attached to the overextended node by one line
\cite{olivegw02}.  The rank  of very extended exceeds by three the rank  
  of the
finite-dimensional simple Lie algebra
${\cal G^{}}$ from which one started.  Some of the properties of very  
extended
algebras including their roots and weights are given in  
\cite{olivegw02}. Thus
M-theory, the closed bosonic string and gravity are associated with   
the very
extension of $E_8$, and of the $D$ and
$A$ series of  finite simple Lie algebras respectively.
\par
It is tempting to suppose that there are also theories associated with  
the other
finite-dimensional simple Lie algebras and that they possess symmetries
that are the corresponding triple extensions ${\cal G^{+++}}$ of these  
algebras.
This more general conjecture is suggested by   dimensional reduction  
and by the
cosmological billiards \cite{damourhn00}.  Given a theory
consisting of gravity coupled   to a dilaton
and a set of forms it will, upon dimensional reduction on a torus, lead  
to a theory
containing scalar fields as well as gauge fields. In general the  
scalars will not
belong to a non-linear realisation. However,  if  one starts with   a   
specific set of
forms and dilaton  coupling to the forms,  one finds that the  scalar  
fields
do belong to a non-linear realisation. When this occurs, and the theory  
is reduced
to three dimensions, all the fields
  can be reduced to scalars which belong to a coset space
${\cal G/H}$ where
${\cal H}$ is the maximal compact subgroup of ${\cal G}$.  In this  
procedure some
of the forms may remain  after dimensional reduction,  but these may be
swapped for scalars   by dualisation.  In fact one can find  all of the
finite-dimensional simple Lie algebras in this way \cite{cremmerjlp99}.  
It
has become customary to refer to the  unreduced theory from which one  
starts
as the oxidation of the theory in  three dimensions.  The most  
important example
is  the 11-dimensional supergravity which can be thought of as a  
maximal
oxidation of the three dimensional theory which has 128 scalar fields   
belonging
to the
$E_8/SO(16)$ coset \cite{ms}.  The oxidation of the three dimensional
theory that   has
coset symmetry $D_n$  is a theory in $n+2$ dimensions that  consists of  
gravity
coupled to a scalar field and a three-form  field strength, the  
couplings of the
fields being fixed.  Reference \cite{lambertw01} considered the   
dimensional
reduction of certain  theories of  gravity coupled to a  dilaton and  
certain
$n$-forms and found that only for very specific couplings did the  
scalars belong
to cosets and in all cases these couplings  were  one of the oxidised  
theories.
The work on  cosmological billiards referred to above was also extended  
to  the
oxidised theory for any ${\cal G}$ and it was found that the one  
dimensional
motion took place in the Weyl chamber of  the overextended algebras
${\cal G^{++}}$ \cite{damourbhs02}.

The main results in this paper are as follows.  Firstly, we have discovered
sets of   solutions of Einstein's equations coupled to scalar fields 
which form linear   representations
of the Weyl  and outer automorphism group  of   $\G$ for {\it any} $\cal G$
and   secondly, we
have shown that $T$ and $S$-dualities in  superstrings and in the  
bosonic string
are encoded in suitable algebras $\G$  in a way which depends   
specifically on the
enlargement of
$\cal G$ up to   their triple Kac-Moody extension.\par

The  content of the paper 
is   organised
as follows. In Section~2, we lay out the general formalism used to  
uncover and
test $\G$ symmetries.  We consider  the Cartan subalgebra  of ${\cal  
G^{+++}}$
with space-time fields as parameters. We identify the latter with the  
diagonal
components of the metric and possible scalar dilaton-like fields.  We   
find  the
invariant metric in the space of the fields and the effect of the Weyl
transformations of ${\cal G^{+++}}$ on  these fields. It is important  
to realise that
although the Cartan subalgebra only contains commuting elements,  it   
allows one
to probe significant aspects of the ${\cal G^{+++}}$ algebra.  One can  
think of this
calculation as carrying out the full non-linear realisation for
${\cal G^{+++}}$,  but then  setting to zero all the fields except  
those in the
Cartan subalgebra.  In Section~3, we show how  the properties of  
dimensional
reduction of `maximally oxidised' theories can be understood in terms  
of the
general framework laid out in Section~2.  We first recall how  reducing  
  these
theories to three dimensions yields the symmetry of the simple Lie  
groups $\cal
G$. We then show how these symmetries can be embedded in an algebra  
${\cal
G^{+++}}$  by using both  the metric of the compactification torus and   
of the
non-compact dimensions as parameters of the algebra. In this way, we  
relate the
invariant metrics and the Weyl transformations of $\cal G$ to those of
${\cal G^{+++}}$. We then formulate a general theory of such Weyl  
preserving
embeddings, which is to be used in the next two sections. In Section~4,  
  we study
solutions of the gravity sector of all maximally oxidised theories. We  
focus our
attention on the Kasner solutions  of gravity with dilatons  and show  
how the set
of these solutions can be enlarged to form representations of the group  
$S(\G)$
of Weyl transformations and  outer automorphisms of   ${\cal G^{+++}}$.  
The
connection of these results with the cosmological billiards is  
explained.
Section~5 discusses  how the group $S(\G)$ encodes the $T$ and
$S$-dualities of strings  for those  oxidised theories which can be  
interpreted as
low energy effective actions of M-theory or string theories. When ${\cal
G^{+++}}$ is restricted to its subgroups acting in the compact  
dimensions, one
recovers well-known results. The present approach however points to more
general links between string theories  and gravity and in particular  
reveals the
precise relation between an $S$-duality of the bosonic string and the  
$S$-duality
in the heterotic string. More importantly perhaps, it shows the full  
power of ${\cal
G^{+++}}$ when the non-compact dimensions are taken into account.  
Namely the
rescaling of the Minkowskian metric due to the transformations of the  
Planck
length under string dualities is automatically taken care of by ${\cal  
G^{+++}}$,
providing a specific signature of this algebra. The significance of  
these results is
discussed in Section~6. 
\par
We now give a systematic account of symmetries in supergravity and
string theories form which two essential elements emerge. The  
first one is the appearance of  non-linear realisations of group  
symmetries in supergravity theories and the second crucial element is
the idea that symmetries, originally   discovered  in
what turned out to be dimensionally reduced theories, are really  
symmetries of
full uncompactified theories. 
\par
Eleven dimensional supergravity has no scalars, and apart from the  
dilaton
in IIA and the two scalar fields  in  IIB,  the supergravity theories  
in ten
dimensions possess  no scalars either.  In lower dimensions however,  
many
  scalars appear and are  found to occur in non-linear realisations, as a
consequence of supersymmetry.   In such constructions,   the scalars     
  belong to
cosets. This feature was  first observed  in the four dimensional
$N=4$ supergravity theory which possesses an   $SL(2,R)$ symmetry  with  
$ U(1)$
as local subgroup, i.e. the scalars belong to the coset
$SL(2,R)/U(1)$~\cite{ferrarasz77}.  One of the most  intriguing  
discoveries in this
context is that       the  maximal supergravity theory in four  
dimensions contains
70 scalars and has
$E_7$ symmetry \cite{cremmerj78}. In fact,  when  11-dimensional  
supergravity
is dimensionally reduced on a $k$-torus to $11-k$ dimensions for
$k=1,\ldots, 8$, the resulting scalars can be expressed as a  non-linear
realisation \cite{julia81} of the sequence of the finite-dimensional   
Lie algebras
$E_k$. The role of GL(D) symmetries and dualities and their relationship 
with the Dynkin diagram of $E_k$ was discussed in \cite{julia98}.
Dimensional reduction of 11-dimensional supergravity to three dimensions
leads to a theory which is invariant under $E_8$
\cite{ms},  to two
   dimensions  leads to   a theory  that is   
invariant   under the
affine extension of $E_8$ \cite{nicolai87}, which is also called $E_9$
(and denoted in   this paper by
$E_8{}^{+}$), while
dimensional reduction to one dimension is thought to result
in a theory that is invariant under   the hyperbolic algebra $E_{10}$
(labelled hereafter   as the
overextension of $E_8$ and denoted by $E_8{}^{++}$)
\cite{julia84}.   In fact the  two scalars in the IIB supergravity   
theory belong to
a $SL(2,R)/ U(1)$ non-linear realisation \cite{schwarzw83} and as the  
dimensional
reduction of this theory on a $k$-torus  agrees with that of the  IIA  
supergravity
theory, one finds the same set of cosets as for this latter theory.

For many years it has been known that any string theory reduced on a   
torus is
invariant under  $T$-duality \cite{kikkaway84}.  Indeed, it has been  
shown
\cite{rocekv92} that this is a  symmetry to all orders of perturbation  
theory. It
has also been conjectured that some string theories possess  
non-perturbative
symmetries called $S$-dualities.  The first such  conjecture was about  
an
$SL(2,Z)$ symmetry in the context of the heterotic string reduced on a   
6-torus
\cite{fontilq90}.   It is believed that,  when restricted to be defined  
on the
integers, all the coset symmetries based on the finite-dimensional  
semi-simple
Lie algebras discussed above  are symmetries of string theory,  called
$U$-dualities \cite{hullt94}. It has also been shown
\cite{elitzurgkr97,obersp98,banksfm98} that the
$T$-duality  transformations for the IIA string theory  in ten  
dimensions reduced
on an $k$-torus for $k=1,\dots,10$  have a natural action on the moduli  
space of
the
$k$-torus that is   the Weyl group of
$E_k$.   It has also been suggested  that  the closed bosonic string
reduced on the torus associated with  the unique self-dual twenty-six
dimensional Lorentzian lattice is invariant under the Borcherds fake  
monster
algebra \cite{moore93} and  there is some evidence of Kac-Moody  or
possibly Borcherds  structures in
threshold corrections of the heterotic string reduced on a six  
dimensional torus
\cite{harveym95}.

  Until recently it was not thought  that the exceptional symmetries  
found in the
dimensional reduction of  11-dimensional supergravity were symmetries of
11-dimensional supergravity itself.   Some time ago however, it was  
shown
\cite{dewitn86} that  11-dimensional supergravity  does possess an
SO(1,2)$\times$SO(16) symmetry, although the SO(1,10) tangent space
symmetry is no longer apparent in this formulation.  It has also been  
noticed
that some of the objects associated with the exceptional groups  
emerging from
the reductions  appear naturally in the unreduced theory  
\cite{meloshn97}.

The coset construction for the scalars was extended \cite{cremmerjlp98}  
to
include the gauge fields of supergravity theories. This method used  
generators
that were inert under Lorentz transformations and, as such, it is  
difficult to
extend the method further to include either gravity or the fermions.   
However,
this construction  did include the  gauge and scalar fields as well as   
their duals,
and as a consequence,  the  equations of motion for these fields could  
be
expressed as a generalised self-duality condition.

More recently, it has been found that the   dynamics of theories of  
gravity
coupled to a dilaton and
$n$-forms near a space-like singularity becomes   a one dimensional  
motion with
scattering taking place in the Weyl chambers of certain overextended Lie
algebras \cite{damourhn00, damourh00}. This motion is sometimes  
referred to as
cosmological billiards.  Supergravity  in ten and eleven dimensions are  
such
theories. One finds that  11-dimensional supergravity, IIA and IIB  
supergravities
all  lead to the same Weyl chamber which is that of  overextended  
$E_8{}^{++}$,
while type I and heterotic supergravities lead to the Weyl chamber of
$B_8^{++}$ \cite{damourh00}.  For gravity alone in D dimensions, the
corresponding Weyl chamber is that of overextended
$A_{D-2}^{++}$ \cite{damourh00,damourhjn01}.

Despite these observations it  has been widely  thought that the  
exceptional
groups found in the dimensional reductions of the maximal supergravity  
theories
can not be  symmetries of 11-dimensional supergravity and must arise as  
a
consequence of the dimensional reduction procedure.  This is  perhaps
understandable given that the non-linear realisations or coset   
symmetries were
associated with scalar fields and there are no scalar fields in  
11-dimensional
supergravity and the two scalars in the IIB theory only belong to the  
non-linear
realisation based on $SL(2,R)/U(1)$. However,
  it has been conjectured
\cite{west01} that  the 11-dimensional  supergravity theory possesses a  
hidden
$E_{11}$ symmetry. This conjecture is based essentially on  two   
results.  Firstly,
it had previously been shown \cite{west00} that the whole
  bosonic sector of 11-dimensional supergravity, including its  
gravitational
sector, was a non-linear realisation. This placed the other bosonic  
fields of the
theory on an equal footing with scalars and showed that non-linear  
realisation of
symmetries can take place, whether or not there are scalars in the  
theory.
Secondly, this construction contained substantial fragments of larger
symmetries including $A_{10}$ and the Borel subgroup of  
$E_7$~\cite{west01}.
This,  and other features of the   non-linear realisation, suggested  
that these
algebras should be incorporated into a Kac-Moody algebra, or in a more  
general
symmetry. Although it was not proven that such a symmetry was realised,  
it was
shown that  if  the symmetry of the theory included a  Kac-Moody algebra
  then it must contain the  very extended
  algebra of $E_8$, called
$E_{11}$~\cite{west01}. In this paper we shall write this triple  
extension of
$E_8$  as
$E_8^{+++}$.  It was also  shown that this construction could be  
generalised to the
ten dimensional IIA \cite{west01} and IIB \cite{schnakw01} supergravity  
theories
and the corresponding Kac-Moody algebra was also
$E_8^{+++}$ in each case.

These ideas were taken up in reference \cite{damourhn02} which    
considered
11-dimensional supergravity as a non-linear realisation of  the
$E_{10}$  subalgebra of
$E_{11}$ in the small tension limit which played a crucial role in  
the work of
references \cite{damourh00,damourhjn01}. These authors also introduced   
  a new
concept of level in the context of
$E_{10}$ which allowed them to deduce the representation content of
$E_{10}$ in terms of representations of
$A_9$ at low levels. They showed that the 11-dimensional supergravity
equations were  $E_{10}$-invariant,  up to level three,
  in the small tension limit,  provided one adopted a particular map  
relating the
fields at a given spatial point to quantities dependent only on time.   
In this limit
the spatial dependence of the fields was very restricted. The $U$-duality
groups   has been 
enlarged into some super Borcherds algebras~\cite{henryjp02}.

The approach given in reference \cite{west01} for M-theory was also  
applied to
the closed bosonic string in twenty-six dimensions. Although this  
theory has no
supersymmetry it is essentially unique with a corresponding low energy
effective action. It was shown
  that this action, generalised to any dimension \cite{west00},  was a  
non-linear
realisation and on similar grounds it was argued \cite{west01} that in  
twenty-six
dimensions it  might possess a symmetry based on $K_{27}$ or very  
extended
algebra of
$D_{24}$, denoted in this paper as
$D_{24}^{+++}$.  It was suggested that it was  this symmetry that is  
responsible
for the uniqueness and spectacular properties of this theory. This is  
consistent
with the old idea \cite{casherent85} that the superstrings in ten  
dimensions are
contained in the closed bosonic string. More recently this work has been
extended to include the branes of the two theories \cite{englertht01}.   
In a
similar spirit it has been conjectured \cite{lambertw01} that  gravity  
in D
dimensions may possess  a Kac-Moody symmetry  which is
$A_{D-3}^{+++}$.

\setcounter{equation}{0}
\section{Group theory of abelian configurations} As explained in the  
introduction
there are indications
\cite{west01}  pointing towards the existence of  a symmetry which
could be  $E_8^{+++}$ in the M-theory approach to a unified theory of  
gravity and
matter. Similarly,  the symmetry  $D_{24}^{+++}$  is thought to  
underlie the
bosonic string theory \cite{west01}.   It has also been suggested that
$A_{D-3}^{+++}$ could be a symmetry of gravity \cite{lambertw01}.   
Sections~4
and 5 will provide explicit examples which not only corroborate these
conjectures but, remarkably, provide evidence for the physical  
relevance of the
triple extension
${\cal G}^{+++}$ of any simple Lie algebra ${\cal G}$.

In this section, we associate fields to the algebra $\G$ in order to  
test these
symmetries. To avoid the complexity of a full non-linear realisation of  
the
algebra, we restrict our attention to its Cartan subalgebra.   We will  
see that this
corresponds to consider  the diagonal components of the metric field and
possibly scalar `dilaton'  fields.   We then calculate the effect of  
the Weyl
transformations and outer automorphism of ${\cal G}^{+++}$  on these
space-time  fields. This  provides a tool for testing hidden symmetries  
in
subsequent sections.

\subsection{General formalism and the preferred subgroup}

Any  finite-dimensional simple Lie algebra
${\cal G}$ can be  extended in a unique way to a Kac-Moody algebra
${\cal G}^{+++}$. This procedure increases the rank of the algebra by  
three and we
refer the reader to reference \cite{olivegw02} for a discussion of this  
process.
The Dynkin diagrams of all the simple algebras ${\cal G}^{+++}$ are  
depicted in
Fig.1 and 2.

Let us denote the rank of ${\cal G}^{+++}$ by $r$ and let  $E_m,F_m$ and
$H_m,
\ m=1,2,\dots r$~, be its Chevalley generators satisfying
\begin{eqnarray}
  [H_m,H_n]=0~~, ~~[H_m,E_n]&=&A_{mn}E_n \, ,\nonumber \\
\label{chevalley}
  [H_m,F_n]&=&-A_{mn} F_n~~,~~[E_m,F_n]=\delta_{mn} H_m\, ,
\end{eqnarray}  where $A_{mn}$ is the Cartan matrix. The Cartan  
subalgebra is
generated by $H_m$, while  the positive (negative) generators  are   the
commutators of the
$E_m$ ($F_n$) subject to the Serre relations
\begin{equation} [E_m,[E_m,\dots,[E_m,E_n]]]=0,\quad
[F_m,[F_m,\dots,[F_m,F_n]]]=0\, ,
\end{equation} where the number of $E_m$ ($F_m$) acting on $E_n$  
($F_n$) is
given by
$1-A_{mn}$. There exists a scalar product  $\langle\, ,\rangle$ on an
$r$-dimensional vector space such that the   Cartan matrix can be  
expressed in
terms of
  the simple roots as
\begin{equation}
\label{cartan}  
A_{mn}=2\frac{\langle\alpha_m,\alpha_n\rangle}{\langle\alpha_m,
\alpha_m \rangle}\, .
\end{equation}

Any given algebra
${\cal G}^{+++}$ contains a preferred
  subalgebra\footnote{Throughout the paper we use the same notation for
groups and algebras.} $GL(D)$ with
$D\le r$ where
$D$ will be identified with the number of space-time dimensions through  
the
introduction of metric fields.  The dimension $D$ is  thus encoded in
${\cal G}^{+++}$,  as will now be  explained.

The generators of $GL(D)$ are taken to be $K^a{}_b,\ a,b=1,2,\ldots ,D$  
with
commutation relations
\begin{equation}
\label{Kcom} [K^a_{~b},K^c_{~d}]  
=\delta^c_{~b}K^a_{~d}-\delta^a_{~d}K^c_{~b}\,  .
\end{equation} The corresponding group elements are
\begin{equation}
\label{ingroup}
\label{group} g=e^{\, h^b_{~a}(x^\mu)K^a_{~b}}\, ,
\end{equation} where the $x^\mu$ are coordinates on a $D$-dimensional  
manifold.
We shall express the  group parameters $ h^b_{~a}(x^\mu)$ in terms of  
the
metric tensor $g_{\mu\nu}=\eta_{ab} e^a_{~\mu} e^b_{~\nu} =\vec  
e_a.\vec e_b
\, e^a_{~\mu} e^b_{~\nu}
$ at the point $x^\mu$.  The vectors $\vec e_a$ form an orthonormal  
basis in the
tangent plane at the point $x^\mu$.  The group generators $K^a_{~b}$  
act on any
vector $\vec z$ in the plane according to
\begin{equation} K^a_{~b}=z^a{\partial\over\partial z^b}\, .
\end{equation}
  Thus, the group element Eq.(\ref{group}) generates the linear  
transformation
\begin{equation}\label{linear} z^{\prime c}=g~ z^c =e^{\,
h^b_{~a}(x^{\mu})z^a\partial/\partial z^b} z^c = \left(
e^{h(x^{\mu})}\right)^c_{~a} z^a\, ,
\end{equation} where $h$ is the matrix whose  elements are the fields $
h^b_{~a}$.   Transforming the basis $\vec e_a$ to $\vec  
e_\mu=e^a_{~\mu}\vec
e_a$ one gets
\begin{equation}
\label{gpara}
  e^a_{~\mu}(x^\mu)=\left( e^{h(x^\mu)}\right)^a_{~\mu}.
\end{equation}  The vielbein group
$SO(D)$ makes it evident that the metric fields  $g_{\mu\nu}(x)$ live  
in the coset
$GL(D)/SO(1,D-1)$.

A more elaborate derivation of Eq.(\ref{gpara}) yielding a
  covariant coupling of the metric to other fields is obtained using  a  
non-linear
realisation of
$GL(D)$ with local subgroup $SO(1,D-1)$. One first introduces the  
extension
$IGL(D)$ of the group $GL(D)$  by adding $D$ momentum generators $P_a$  
to its
generators
$K^a{}_b$ and carries out simultaneously the  non-linear realisation  
with the
conformal group \cite{borisovo74,west00}.

The selection of the preferred subalgebra $GL(D)$ of ${\cal G}^{+++}$  
is achieved
as follows.   Starting from the  root of the Dynkin diagram  that  
extends ${\cal
G}$ up to
${\cal G}^{+++}$, labelled 1 in Fig.1 and Fig.2,  one follows the line  
of long roots up
to the last one
  whose Chevalley generator has the form $H_m=\delta _m^a
(K^a{}_a-K^{a+1}{}_{a+1})$. This line constitutes the Dynkin diagram of  
an
$A_{D-1}$ subgroup and is referred to  as the gravity line{\footnote {In case
there is a bifurcation in the Dynkin diagram, we select the line consistent with
the $GL(D-3)$ subgroup of theory in three dimensions that results from the
corresponding  maximally oxidised theory. }}.

\hskip 1cm \epsfbox{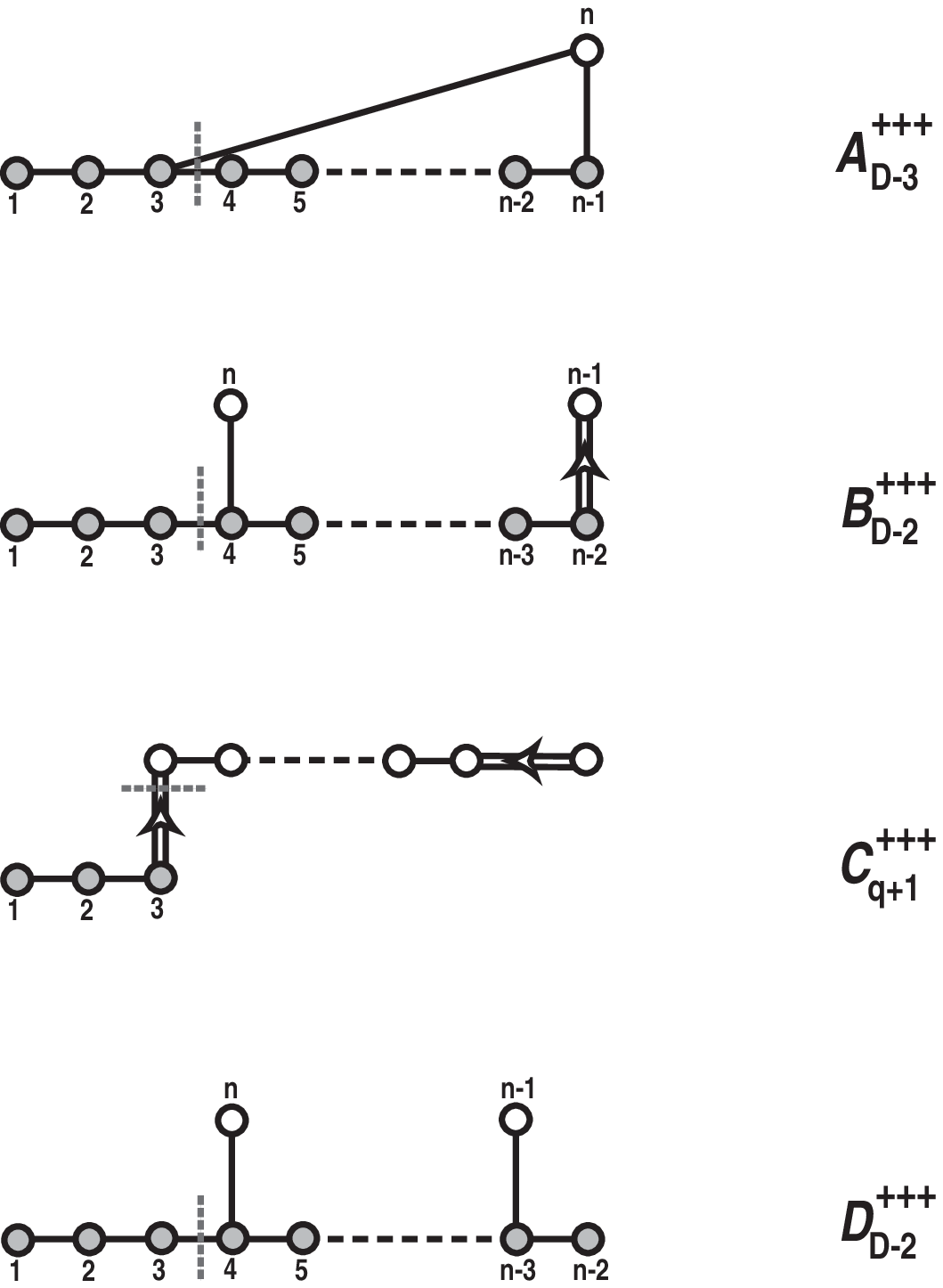}

\vskip1.5cm
\begin{quote}\begin{center}
\baselineskip 14pt {\small Fig.1.  Dynkin diagram of $\G : A, B, C, D$  
series.}
\end{center} {\small The nodes of the gravity line are shaded. The  
nodes labelled
1,2,3 are the Kac-Moody extensions of the  Lie algebras.}
\end{quote}

\hskip 2cm \epsfbox{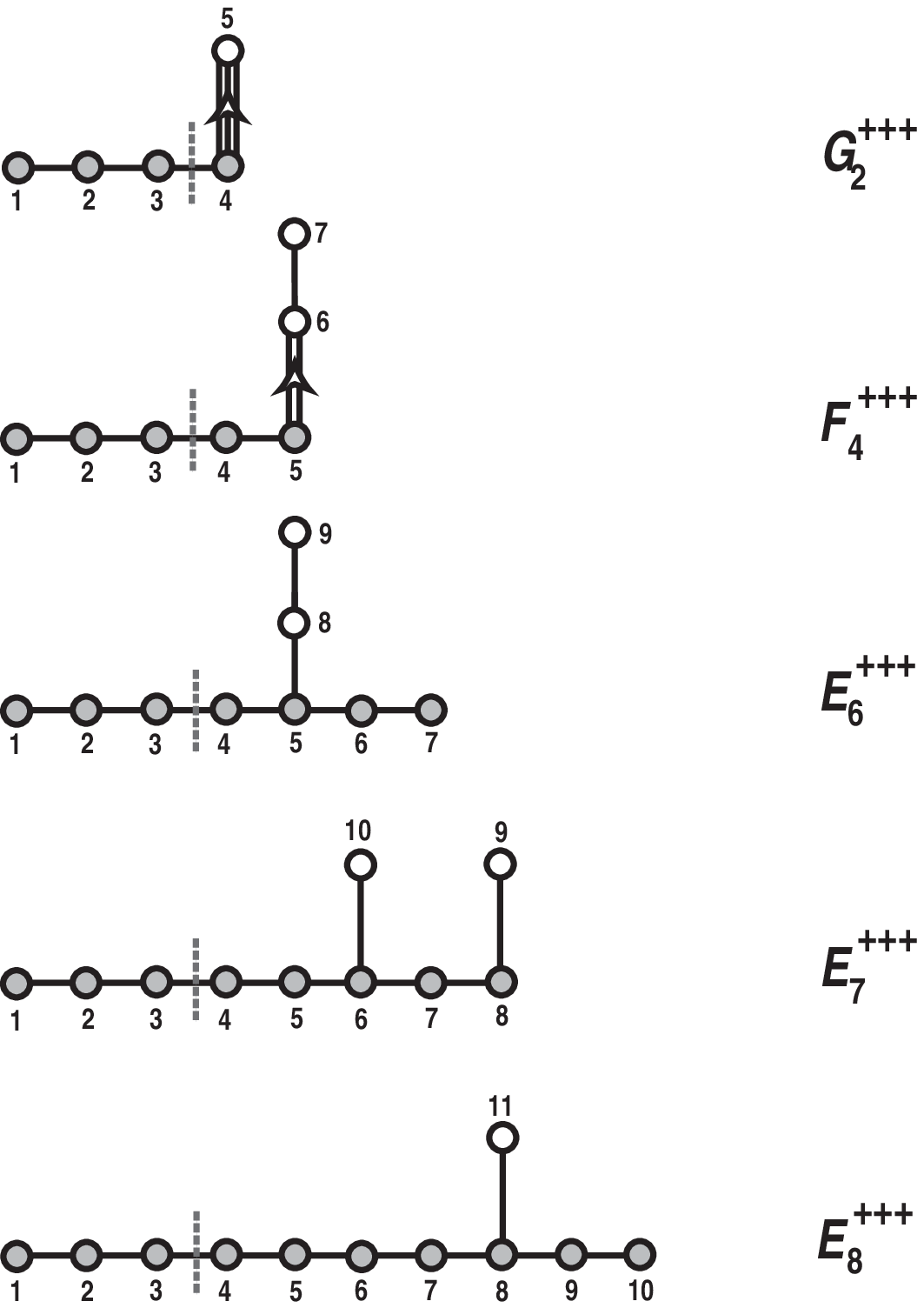}

\begin{quote}\begin{center}
\baselineskip 14pt {\small Fig.2.  Dynkin diagram of $\G$ : exceptional  
algebras.}
\end{center}
\end{quote} 

The above  
expressions
for the $H_m$ generators are  identified by noting that the simple  
positive roots
of $SL(D)$ are $K^a{}_{a+1},\ a=1,\ldots ,D-1$. The additional commuting
generator  $\sum_{a=1}^D K^a{}_a$ extends $SL(D)$ to the preferred   
$GL(D)$. The rank $r$ of  some of the algebras ${\cal G}^{+++}$ is  
equal to the
rank $D$ of the preferred $GL(D)$ subalgebra.  In these cases the  
embedding of
$GL(D)$ is such that  the Cartan  generators $H_m$ of ${\cal G}^{+++}$   
in the
Chevalley basis can be expressed in terms of  the generators
$K^a{}_a, a=1,\ldots ,D$ and vice versa. For some ${\cal G}^{+++}$  
algebras
however,
$D<r$ and  we enlarge the  preferred subalgebra to
\begin{equation} GL(D)\oplus R_1\oplus\dots \oplus R_q\, ,
\end{equation}
  where $[R_u, K^a{}_b]=0,\ u=1,\ldots ,q$. The  number of extra  
commuting
generators $R_u$ is $q=r-D$.  These additional generators  will be   
associated
with  scalar `dilaton'  fields
$\Phi^u$.  The occurrence of such fields will be corroborated by the  
analysis of
Section~3.  For each simple
${\cal G}^{+++}$, the number
$q$ of dilatons  can be read off its Dynkin diagram and is given in the  
Table
I.

\begin{center}
\begin{tabular}{||c|c|c||}
\hline $q$&$D$&$ {\cal G}^{+++}$\\
\hline\hline
$0$&$D$&$A_{D-3}^{+++}$\\
\hline
$1$&$D$& $B_{D-2}^{+++}~{\rm and}~ D_{D-2}^{+++}$\\
\hline
$q$&$4$& $C_{q+1}^{+++}$\\
\hline
$1$&$8$& $E_6^{+++}$\\
\hline
$1$&$9$& $E_7^{+++}$\\
\hline
$0$&$11$& $E_8^{+++}$\\
\hline
$0$&$5$& $G_2^{+++}$\\
\hline
$1$&$6$& $F_4^{+++}$\\
\hline
\end{tabular}
\vskip .5cm
   Table I
\end{center}

We now restrict ourselves to the Cartan subalgebra of ${\cal G}^{+++}$  
which we
express in terms of the generators associated with gravity and scalar  
dilaton
fields, namely  we consider the $L$-basis defined by
   \begin{equation}
    \label{Kbasis} L_i=\{K^a_{~a}, R_u\}\qquad i=1,2,\dots, q+D=r\, .
   \end{equation} The corresponding abelian group element $g$ is
\begin{equation}
\label{explicit} g=\exp (\sum_{a=1}^D p^a K^a{}_a) \exp (\sum_{u=1}^q  
\Phi^u
R_u)\, .
\end{equation} Comparing Eq.(\ref{explicit}) with the group element  
given in
Eq.(\ref{ingroup}) and using Eq.(\ref{gpara}), we find that taking a   
group element
restricted to the Cartan subalgebra corresponds in the gravity sector  
to  taking
a vielbein that is diagonal. The coordinates
$p^a$ are thus related to a diagonal metric  configuration  
by\footnote{The
indices in $g_{aa}$ are identified by the correspondence  between $p^a$  
and
$e^a_{~\mu}$ in a given  frame.}
\begin{equation}
\label{metriccoord} e^a_{~\mu}=\left(
e^h\right)^a_{~\mu}=e^{p^a}\delta_\mu^a~~~ {\rm or}~~~  g_{aa}=
e^{2p^a}\eta_{aa}\, .
\end{equation} As already mentioned,  the additional $r-D$ coordinates  
are
  the dilaton fields. The group element Eq.(\ref{explicit}) can also be  
written in
the Chevalley basis as
\begin{equation}
   \label{qpara} g=e^{\, q^m H_m}\, ,
\end{equation} and the $H_m$ are related to the $L_i$ by a linear  
relation
  \begin{equation}
    \label{rootK} H_m=r_m^{~i}\,  L_i\, .
    \end{equation}

Given any symmetrisable  Kac-Moody algebra there  exists, up to a
numerical factor, a   unique
scalar product defined on the algebra that is invariant under the  
adjoint action of
the algebra \cite{kac83}.   For a finite-dimensional simple Lie algebra  
this
is just the Killing form which can be expressed as the trace of the  
generators in
any finite-dimensional representation. While the expression for the  
scalar
product   on the full Kac-Moody algebra may be complicated, its  
expression on
the Cartan subalgebra, up to a numerical scale factor,    is given in  
terms of the
Cartan matrix by \cite{kac83}
\begin{equation}
   \label{groupmetric}  (G_c)_{mn}={2A_{mn}\over\langle\alpha_n,\alpha_n
\rangle}=\frac{4\langle\alpha_m,\alpha_n\rangle}  
{\langle\alpha_m,\alpha_m
\rangle\langle\alpha_n,\alpha_n
\rangle}\, .
    \end{equation} This  invariant metric  will be used in deriving Weyl  
and other
symmetries in theories containing gravity and extra degrees of freedom.  
    We
now show how to obtain the invariant metric in the space of the  
physical fields
defined in the $L$-basis.

We have
   \begin{equation}
\label{HK} g=e^{\, q^m H_m}=e^{\, p^i L_i} \qquad {\rm where} \quad  
p^i= q^m
r_m^{~i}
\quad{\rm or}\quad{\bf p^T= q^T\cdot r} \, ,
\end{equation} and  the $p^i$ are
\begin{equation}
\label{Kcord} p^i=\{p^a, \Phi^u\}\, ,
\end{equation} with $p^a$ given in Eq.(\ref{metriccoord}). In terms of  
the
$L$-basis the invariant metric is
   \begin{equation}
   \label{modmetric}
      G^{+++}_{ij} = [{\bf r^{-1} \cdot G_c \cdot (r^{-1})^T}]_{ij}\, ,
\end{equation} where  $\bf {G_c}$ is  given in Eq.(\ref{groupmetric}).

We shall see  in the examples below and in Section~3 that $  
G_{ij}^{+++}$ has the
same form for {\em all} the algebras
${\cal G}^{+++}$  and  is equal to
\begin{equation}
\label{very} {\bf G^{+++}}=\frac{1}{2}{\bf I}_{r-D} \oplus ({\bf I}_D -
\frac{1}{2}{\bf\Xi}_D)\, ,
\end{equation} where ${\bf\Xi}_D$ is a $D$-dimensional matrix  with all  
entries
equal to one. The reason behind this universality is that   
Eq.(\ref{modmetric}) can
be formally written as
\begin{equation}
\label{trace}
   G^{+++}_{ij} = ({\bf r^{-1}})_i^{~m}\, \langle H_m,H_n\rangle\, ({\bf
r^{-1}})_j^{~n}=\langle L_i,L_j\rangle\, ,
\end{equation} and that the last expression involves only generators of  
the
preferred subgroup.  The part of the scalar product involving the  
generators of
$SL(D)$ is fixed as it must be the unique, up to a scale, invariant  
scalar product on
this algebra.  We have used  the arbitrariness of the normalisation of  
the dilaton
fields to fix the factor multiplying ${\bf I}_{r-D}$ in the metric  
Eq.(\ref{very}) to
be $1/2$. Note that we can also write Eq.(\ref{modmetric}) as
\begin{equation}
\label{rootG} ({\bf r})_m^{~i}G^{+++}_{ij}({\bf r})_n^{~j} =
(G_c)_{mn}=\langle\alpha_m^\vee,\alpha_n^\vee\rangle\, ,
\end{equation} where the coroots $\alpha_n^\vee$ are
\begin{equation}
\label {coroot}
   \alpha_n^\vee=\frac{2\alpha_n} {\langle\alpha_n,\alpha_n
\rangle}\, .
\end{equation}
  Eq.(\ref{rootG}) identifies the matrix of the contravariant components  
of the
coroots in the $L$-basis with the matrix ${\bf r}$.

We now show how to use our formalism to obtain the action of the ${\cal  
G}^{+++}$
Weyl transformations   on the fields.

The Weyl transformations $S_\alpha $ of any Kac-Moody algebra are  
reflections
in the planes perpendicular to the real roots $\alpha$ of the algebra.
Their   action on
an arbitrary weight $\omega$ is given by
\begin{equation}
\label{weylgen}
\omega' \stackrel{def}{=} S_\alpha\, \omega = \omega
-2\frac{\langle\omega,\alpha\rangle}{\langle\alpha,\alpha
\rangle}\alpha\, .
\end{equation} The Weyl group is  generated by the reflections  
associated with
the simple roots $\alpha_n$.  Their action  on the  simple coroots
$\alpha_m^\vee$ is
\begin{equation}
   \label{weylsimple}
\alpha_m^{\vee \prime} \stackrel{def}{=}
S_{\alpha_n}\alpha_m^\vee=(s_{\alpha_n})_m^{~p}\alpha_p^\vee=\alpha_m^\vee-A_{mn}
\,\alpha_n^\vee\, .
   \end{equation} Since they are reflections, ${\bf s}_{\alpha_a}^2={\bf  
1}$.
Although the  full Kac-Moody algebra does not carry a representation of  
the
Weyl group, the Cartan subalgebra does, and it takes the form
   \begin{equation}
    \label{weylcartan} H_m^\prime
\stackrel{def}{=}S_{\alpha_n}H_m=(s_{\alpha_n})_m^{~p}H_p\,   .
    \end{equation} The corresponding action on the fields $q^m$ in the  
group
element Eq.(\ref{qpara})  is  given by
   \begin{equation}
\label{weylpara} {\bf q^{\prime T}}={ \bf q^T\cdot  s}_{\alpha_n}\, .
\end{equation} The invariant scalar product on any Kac-Moody algebra,  
discussed
above, is also invariant under Weyl  reflections. When restricted to  
the Cartan
subalgebra elements this can be seen  as consequence of the fact that
reflections preserve the  scalar products of the roots. Consequently  
the metric
$(G_c)_{mn}$ of equation Eq.(\ref{groupmetric}) is invariant under the  
Weyl
group, i.e
${\bf s}_{\alpha_n}{\bf \cdot\, G_c\cdot s}_{\alpha_n}^{\bf  T}{\bf   
=G_c}$ for all
simple roots ${\alpha_n}$.
   Using the relation Eq.(\ref{HK}) we find that the generators of the  
Weyl group
acting on the physical fields
$p^i$ are given by
   \begin{equation}
   \label{pweyl} {\bf {p^\prime}^T = p^T \cdot t}_{\alpha_n}\, ,
    \end{equation} where
\begin{equation}
    \label{weyl} {\bf   t}_{\alpha_n}{\bf= r^{-1}}{\bf \cdot\,   
s}_{\alpha_n}{\bf
\cdot r}\, .
    \end{equation} The  corresponding action of the Weyl group on  the  
generators
$L_i$ is thus given by  $L_i' =(t_{\alpha_n})_i{}^j L_j$ and the  
invariance of
   the metric $G_{ij}$ is expressed by ${\bf t}_{\alpha_n}{\bf \cdot   
G\cdot
t}_{\alpha_n}^{\bf T}{\bf=G}$. Since we have identified  diagonal
  space-time metrics in equation Eq.(\ref{metriccoord}) in terms of
$p^a$, we can read off  the effect of a Weyl transformation on such  
metrics. For
the algebras that require additional $R_u$ generators we can extend this
procedure to Weyl induced transformations on the diagonal   metric  
components
and on the fields $\Phi^u$. This is one of the central results of this  
section and is
worked out in detail for several important cases below.
\par The above construction can be viewed as a restriction of a  
non-linear
realisation of $\cal G^{+++}$. To see this, let us  consider a  
non-linear
realisation of  $\cal G^{+++}$ where the local
     subgroup is  the subgroup  invariant under the Cartan involution.   
The
generators of the  Cartan subalgebra of $\cal G^{+++}$ are not   
contained in  the
local subgroup as they are not invariant under the Cartan involution.  
As such,
they  lead to the finite set of fields
$q^m$ in the non-linear realisation.
  We can think of the  group elements  Eq.(\ref{qpara})  as those of the  
full
non-linear realisation, but  with all the fields set to zero except for  
those
associated with the Cartan subalgebra. It is important to realise that  
in doing so
we  can capture more information than that  just contained in the Cartan
subalgebra. This occurs in two ways:  we have used the information  
about the
embedding of the physically relevant preferred
$GL(D)\oplus R_1\oplus \ldots \oplus R_q$ in the full $\cal G^{+++}$  
algebra and
we have also  used  a scalar product, or metric
$(G_c)_{mn}$, which is the restriction of the scalar product  invariant  
under the
full $\cal G^{+++}$ algebra.  This  explains why we shall find  results  
relevant to  M
and more general theories from an analysis of their gravity sector.

We now illustrate the above general formalism by applying it to
  several important examples.

\subsection{$E_8^{+++}$ or M-theory}

We now consider the algebra $E_8^{+++}$ which is the suggested symmetry  
of
M-theory \cite{west01}.  The algebra $E_8^{+++}$ admits $SL(11)  
=A_{10}$ as its
gravity line as can be seen on the Dynkin diagram of Fig.2: it is  
obtained by
deleting the node labelled
$n$. Since the rank of $E_8^{+++}$ is eleven, which is also the number  
of
commuting generators $K^a{}_a$ in
$GL(11)$, there are  no additional $R_u$ generators. The relation  
between the
commuting generators $K^a{}_a$ of the preferred
$GL(D)$ and   the Cartan generators $H_m$ of $E_8^{+++}$ in the   
Chevalley basis
follows from comparing the commutation relations Eqs.(\ref{chevalley})  
and
(\ref{Kcom}) and from the identification of the simple roots of
$E_8^{+++}$. These are
$E_m=\delta_m^{a} K^a{}_{a+1},\ m=1,\ldots ,10$ and  $E_{11}= R^{\, 9\,  
10\,11}$
where
$R^{abc}$ is a generator in
$E_8^{+++}$ that is a third rank anti-symmetric tensor under $SL(11)$   
and thus
obeys the equation \cite{west01}
\begin{equation} [K^a{}_b, R^{c_1c_2c_3}]=\delta  
_b^{c_1}R^{ac_2c_3}+\dots\, ,
\end{equation}
   where $+\ldots $ denotes the corresponding anti-symmetrisation. This  
also
follows from the analysis of Section~3. Given this commutator, it is
straightforward to verify that the following relations between $H_m$ and
$K^a{}_a$  are correct\footnote{In what follows, the indices in the  
diagonal
$K^a{}_a$ are considered as a single index as far as summation  
conventions are
concerned.} as they reproduce the Cartan matrix
$A_{mn}$ of
$E_8^{+++}$ using the relation
$[H_m,E_n]=A_{mn}E_n$.
\begin{eqnarray}
\label{aa} H_m&=&\delta_m^{a}(K^a{}_a-K^{a+1}{}_{a+1})~~~~  
m=1,\dots,10\\
\label{el} H_{11}&=&-{1\over 3}(K^1{}_1+\ldots +K^8{}_8) +{2\over  
3}(K^9{}_9+
K^{10}{}_{10}+K^{11}{}_{11})\,.
\end{eqnarray}

Since $H_m=r_m^{~a}\, K^a{}_a$ one readily reads off the matrix ${\bf   
r}$. The
metric $G_{ij}$ in the $L$-basis can therefore be computed using
Eq.(\ref{modmetric}). One gets
\begin{equation}
\label{Mmetric} {\bf G^{+++}}={\bf I}_{11} -
\frac{1}{2}{\bf\Xi}_{11}~~{\rm and \ its\  inverse }~~ {\bf G^{-1}_{  
+++}}={\bf
I}_{11} -
\frac{1}{9}{\bf\Xi}_{11}\, .
\end{equation}
   In calculating the inverse we use the fact that for any $D$ the   
inverse of the
matrix ${\bf I}_D +a\, {\bf\Xi}_D$ is ${\bf I}_D -[a/(1+Da)]\,  
{\bf\Xi}_D$.
\par We now give the effect of the Weyl transformations on the
$p^a$ fields. This is computed via Eq.(\ref{weyl}), or equivalently, in  
the following
way. Let us first consider the Weyl transformation generated by a  
simple root
$\alpha_n,\ n\not=1,8,10,11$. Its action on the Chevalley generators  
$H_m$ is
easily read off  Eqs.(\ref{weylsimple}), (\ref{weylcartan}):
\begin{eqnarray} S_{\alpha_n}H_m&=&H_m\qquad\quad~~ m\ne n,\ m\ne n\pm  
1\,
,\nonumber \\ S_{\alpha_n}H_n&=&-H_n\qquad\quad S_{\alpha_n} H_{n\pm
1}=H_n+H_{n\pm1}\, .
\end{eqnarray} It is then straightforward to see from Eq.(\ref{aa})    
that  the
corresponding transformation on  $K^a{}_a$ for $a=n$ is given by
\begin{equation}
   K^a{}_a\leftrightarrow K^{a+1}{}_{a+1}\, .
\end{equation} This relation is actually valid for $ \ n=1,\ldots 10$,  
as is easily
verified separately for
$n=1,8,10$.  Hence on the fields
$p^a \, (g_{aa}=
e^{2p^a}\eta_{aa}) $, the Weyl  transformation associated to
a simple root
$\alpha_n\, ,\ n=1,\dots ,10\, $ induces the transformation
\begin{equation} p^a\leftrightarrow p^{a+1}\qquad\quad a=n,\,   
n\neq11\, ,
\end{equation} all other $p^a$ being unchanged. In other words each  
such Weyl
transformation just interchanges two of the diagonal components of the  
metric.
\par It remains to find the action of the Weyl transformation generated  
by the
simple root $\alpha_{11}$. In this case
\begin{eqnarray} S_{\alpha_{11}}H_m&=&H_m\qquad\qquad m\ne 8,\
m\ne11\, ,\nonumber\\ S_{\alpha_{11}}H_{11}&=&-H_{11}\qquad\quad
S_{\alpha_{11}} H_{8}=H_8+H_{11}\, .
\end{eqnarray} Examining the $H_m$ that are unchanged we conclude that
\begin{equation} K^a{}_a\to K^a{}_a+y\quad a=1,\ldots ,8\qquad\qquad  
K^a{}_a\to
K^a{}_a+x\quad  a=9,10,11\, .
\end{equation} It is then straightforward to show that $y=0$ and  
$x=-H_{11}$. As
a result, we find that
\begin{eqnarray} S_{\alpha_{11}}K^a{}_a&=&K^a{}_a,\quad a=1,\ldots ,8
\\ S_{\alpha_{11}}K^a{}_{a}&=&K^a{}_a+{1\over
3}(K^1_{~1}+K^2_{~2}+\dots+K^8_{~8})-{2\over
3}(K^{9}_{~9}+K^{10}_{~10}+K^{11}_{~11})~~ a=9,10,11 .\nonumber
\end{eqnarray} Hence we conclude that
\begin{eqnarray}
\label{WeylMe2} p^{\prime a}&=&p^a +\frac{1}{3}(p^9+p^{10}+p^{11})  
\qquad
a=1,\dots, 8\\
\label{WeylMe1} p^{\prime a}&=&p^a -\frac{2}{3}(p^9+p^{10}+p^{11})  
\qquad
a=9,10,11\, .
\end{eqnarray} These relations will be used in Sections~4 and 5 in the  
context of
M-theory whose effective bosonic Lagrangian is \cite{cremmerjs78}
\begin{equation}
    \label{Mth}
   {\cal L}^{(11)} =\sqrt{-g^{(11)}}\left(R^{(11)}- {1\over 2  . 4!
}F^{(3)}_{\mu\nu\sigma\tau}F^{(3)\mu\nu\sigma\tau}+ CS \hbox  
{-term}\right)\, .
\end{equation} One can generalise the above discussion for the  
$E^{(k)},\ k\le 11$
subgroups obtained by putting $q_1,\dots, q_{11-k}=0$. All the above  
discussion
and formulae hold after making the appropriate changes in the labelling  
of
indices. The only exception is the metric Eq.(\ref{Mmetric}) which is  
given by,
(see also reference \cite{banksfm98})
\begin{equation}
\label{partmetric}
  {\bf G^{(k)}}={\bf I}_k +
\frac{1}{9-k}{\bf\Xi}_k \ {\rm and \ its\  inverse \ by }\ {\bf  
G^{-1}_{(k)}}={\bf I}_k
-
\frac{1}{9}{\bf\Xi}_k\, .
\end{equation} We shall come back to this point, and in particular to  
the case
$k=9$, in Section~3.

\subsection{$D_{D-2}^{+++}$ and the bosonic string}

The closed bosonic string in twenty-six dimensions is thought  to  
possess a
$D_{24}^{+++}$ symmetry \cite{west01}. The Dynkin diagram of
$D_{D-2}^{+++}$ is given in Fig.1. The gravity line $SL(D)=A_{D-1}$ is  
obtained by
deleting the two  nodes
  $n\, (=D+1)$ and $n-1$ in the Dynkin diagram. The rank of
$D_{D-2}^{+++}$ is  $D+1$ and we must add one generator $R$ associated  
with the
dilaton, as indicated in Table I.

The simple roots of $D_{D-2}^{+++}$ are
$E_m=\delta_m^{a}K^a{}_{a+1},\ m=1,\ldots ,D-1$,   $E_{n-1}= R^{n-2\,  
n-1}$ and
$E_{n}= R^{5\ldots n-1}$ where
$R^{ab}$ and $R^{a_1\ldots a_{n-5}}$ are generators of
$D_{D-2}^{+++}$ that are  2 and $n-5$ rank anti-symmetric tensors under
$SL(D)$ and thus obey the equations \cite{west01}
\begin{equation} [K^a{}_b, R^{c_1c_2}]=\delta _b^{c_1}R^{ac_2}+\dots\,  
, ~~~
[K^a{}_b, R^{c_1\dots c_{n-5}}]=\delta_b^{c_1}R^{a\dots  
c_{n-5}}+\dots\, ,
\end{equation} as well as
\begin{equation}
\label{normal} [R, R^{c_1c_2}]={l\over2}R^{c_1c_2}\, , ~~~ [R,  
R^{c_1\ldots
c_{n-5}}]=-{l\over 2} R^{c_1\dots c_{n-5}}~\,  ,
\end{equation} with
\begin{equation}
\label{adil} l =\left({8\over D-2}\right)^{1/2}\, .
\end{equation} These also follow from the analysis of Section~3. The
normalisation of the dilaton generator in Eq.(\ref{adil})  is chosen for
convenience. From these commutators, one gets as in the previous case  
the
relations between the generators
$H_m$ and the generators $\{K^a{}_a, R\}$, namely
\begin{eqnarray} H_m&=&\delta_m^{a}(K^a{}_a-K^{a+1}{}_{a+1})\qquad
m=1,\ldots,n-2\\ H_{n-1}&=&-{2\over(D-2)}(K^1{}_1+\ldots
+K^{n-3}{}_{n-3})\nonumber\\ &&+{(D-4)\over (D-2)}(K^{n-2}{}_{n-2}+
K^{n-1}{}_{n-1})+l\, R\, ,\\ H_{n}&=&-{(D-4)\over(D-2)}(K^1{}_1+\ldots
+K^{4}{}_{4})\nonumber\\&& +{2\over (D-2)}(K^{5}{}_{5}+\ldots +
K^{n-1}{}_{n-1})-l\, R\, .
\end{eqnarray}

The metric $G_{ij}$ appropriate to the  $p^i$ fields can be calculated  
using
Eq.(\ref{modmetric}).  One obtains
\begin{equation}
\label{metricD} {\bf G^{+++}}=\frac{1}{2}{\bf I}_1 \oplus ({\bf I}_D -
\frac{1}{2}{\bf\Xi}_D)\, ,
\end{equation} in agreement with the general expression  
Eq.(\ref{very}). Its
inverse is given by
\begin{equation}
\label{invmetricD} {\bf G^{-1}_{ +++}}= 2\, {\bf I}_1 \oplus ({\bf I}_D  
-
\frac{1}{D-2}{\bf\Xi}_D)\, .
\end{equation}

The determination of the Weyl transformations follows the same steps as
  in the previous example. The Weyl transformations corresponding to  
simple
roots in the gravity line  induce an exchange of neighbouring  
components of the
diagonal metric and leave $ \Phi$ invariant.

The Weyl transformation corresponding to  the simple root $\alpha_{n-1}$
induces the changes
\begin{eqnarray} &&K^a{}_a\to K^a{}_a\quad a=1,\dots ,n-3\nonumber\\ &&
K^a{}_a\to K^a{}_a -H_{n-1}\quad a=n-2 ,n-1\qquad\qquad R\to R -{l\over
2}H_{n-1}\, ,
\end{eqnarray} from which we derive
\begin{eqnarray}
\label{Del2} &&p^{\prime a}=p^a+{2\over (D-2)}(p^{n-2}+p^{n-1}) +{l\over
D-2}\Phi\qquad a=1,\dots ,n-3\\
\label{Del1} &&p^{\prime a}=p^a -{(D-4)\over (D-2 )}(p^{n-2}+p^{n-1})
-{l\,(D-4)\over 2\, (D-2)}\Phi\qquad a=n-2,n-1\qquad\\
\label{Del0} &&\Phi^{\prime}=\Phi-l(p^{n-2}+p^{n-1})-{4\over D-2}\Phi\,  
  .
\end{eqnarray}

The Weyl transformation corresponding to  the simple root
$\alpha_{n}$ induces the changes
\begin{eqnarray} &&K^a{}_a\to K^a{}_a\quad a=1,\dots ,4\nonumber\\
&&K^a{}_a\to K^a{}_a -H_{n}\quad a=5,\ldots ,n-1\qquad \qquad R\to R  
+{l\over
2}H_{n}\, .~~~~~~
\end{eqnarray} Hence  $p^a$ and $\Phi$ transform as
\begin{eqnarray}
\label{Dm2} &&p^{\prime a}=p^a+{D-4\over D-2}(p^{5}+\dots +p^{n-1})  
-{l\,
(D-4)\over 2\, (D-2)}\Phi\qquad a=1,\dots ,4\\
\label{Dm1} &&p^{\prime a}=p^a -{2\over D-2}(p^{5}+\ldots +p^{n-1})  
+{l\over
(D-2)}\Phi\qquad a=5,\ldots ,n-1\qquad \\
\label{Dm0} &&\Phi^{\prime}=\Phi+l\, (p^{5}+\ldots +p^{n-1}) -{4\over
(D-2)}\Phi\, .
\end{eqnarray}

Unlike the previous case, $D_{D-2}^{+++}$ possesses an outer  
automorphism
which manifests itself in the symmetry of the Dynkin diagram found by
exchanging nodes $n-2$ and $n-1$. This will also be an invariance of  
the metric
and we can compute its effect on the fields  in the same way as for Weyl
transformations. Under the outer automorphism the $H_m$ transform as
\begin{equation} H_{n-2}^\prime=H_{n-1}\qquad H_{n-1}^\prime=H_{n-2}
\qquad H_m^\prime=H_m\quad m=1,\dots , n-3, n \, .
\end{equation} Hence
\begin{eqnarray} &&K^{\prime n-1}{}_{n-1}=
K^{n-1}{}_{n-1}+H_{n-2}-H_{n-1}\, ,\quad R^\prime= R+{l\over
4}(H_{n-2}-H_{n-1})\, ,\nonumber\\ &&K^{\prime a}{}_a= K^a{}_a\qquad
a=1,\ldots,n-2\,  .
\end{eqnarray} The corresponding change induced on the $p^a$ and $\Phi$  
is given
by
\begin{eqnarray}
\label{Dout2} &&p^{\prime a}=p^a +{2\over (D-2)}p^{n-1}+{l\over 2\,
(D-2)}\Phi\qquad  a=1,\dots  n-2\\
\label{Dout1} &&p^{\prime n-1}=p^{n-1}-{2(D-3)\over (D-2)}p^{n-1} -{l\,
(D-3)\over 2\, (D-2)}\Phi\\
\label{Dout0} &&\Phi^{\prime}=\Phi -{2\over D-2}\Phi -l\, p^{n-1}\, .
\end{eqnarray} These relations will be used in Sections~4 and 5 in the  
context of
the bosonic string theory whose effective  Lagrangian is, for $D=26$,  
given  by
\cite{fradkint85}
\begin{equation}
    \label{Bth}
   {\cal L}^D =\sqrt{-g^D}\left(R^D-
\frac{1}{2}\partial_\mu\Phi\partial_\nu\Phi-{1\over 2 . 3!
}e^{-l\Phi}F^{(3)}_{\mu\nu\sigma}F^{(3)\mu\nu\sigma}\right)\, ,
\end{equation} where $l$ is defined by Eq.(\ref{adil}).

\subsection{$B_{D-2}^{+++}$ and the heterotic string} The Dynkin  
diagram of
$B_{D-2}^{+++}$ is given in Fig.1. The gravity line $SL(D)=A_{D-1}$ is  
obtained by
deleting the two  nodes
  $n\, (=D+1)$ and $n-1$ in the Dynkin diagram, where $n-1$ is  the  
short root. The
rank of
$B_{D-2}^{+++}$ is  $D+1$ and we must again add one generator $R$  
associated
with the dilaton, as indicated in Table I.

Following the previous method, one gets results similar to those for the
$D$-series. The metric for the $p^i$ fields is given by  
Eqs.(\ref{metricD}) and
(\ref{invmetricD}), in agreement with the  universal formula  
Eq.(\ref{very}).
Again, the Weyl transformations corresponding to simple roots in the  
gravity
line  induce an exchange of neighbouring components of the diagonal  
metric and
leave $ \Phi$ invariant. The Weyl transformation for the simple root  
$\alpha_n$
yields the same result as previously, namely Eqs. (\ref{Dm2})-  
(\ref{Dm0}) while
the Weyl transformation generated by the short root $\alpha_{n-1}$  
yields the
transformation Eqs.(\ref{Dout2})-(\ref{Dout0}). Thus we see that the  
outer
automorphism of the
$D$-series is converted in the $B$-series to the Weyl reflection  
corresponding
to  the short root. These relations will be used in Section~5 in the  
context of the
heterotic  string theory whose bosonic effective Lagrangian is, for
$D=10$,
   related to \cite{brinkss77}
\begin{eqnarray}
\label{Hth} {\cal L}^D &=&\sqrt{-g^D}\left(R^D-
\frac{1}{2}\partial_\mu\Phi\partial_\nu\Phi-{1\over 2 . 3! }e^{-
l\Phi}F^{(3)}_{\mu\nu\sigma}F^{(3)\mu\nu\sigma}\right.\nonumber\\
&&\left.-{1\over 2.2! }e^{- (l/2)\Phi}F^{(2)}_{\mu\nu}F^{(2)\mu\nu}+  
C.S.\right)\, ,
\end{eqnarray} where $l$ is given by Eq.(\ref{adil}).

\subsection{$A_{D-3}^{+++}$ or gravity}

The Dynkin diagram of $A_{D-3}^{+++}$ is given in Fig.1.
  The gravity line $SL(D)=A_{D-1}$ is obtained by deleting the node
  $n=D$  in the Dynkin diagram.   The rank of $A_{D-3}^{+++}$  is $D$  
and we need no
further dilaton generator, as indicated in Table I.

The simple roots of $A_{D-3}^{+++}$ are
$E_m=\delta_m^{a}K^a {}_{a+1},\ m=1,\dots ,n-1$ and $E_{n}= R^{4\ldots  
n-1,n}$
where
   $R^{a_1\dots a_{n-4},b}$ is totally anti-symmetric in its $a$ indices  
and the part
antisymmetrised in all its indices  vanishes \cite{lambertw01,west02}.  
This
generator of
$A_{D-3}^{+++}$  transforms under
$SL(D)$ as the indices suggest    and obeys the commutator
\begin{equation} [K^a{}_b, R^{c_1\dots  
c_{n-4},d}]=\delta_b^{c_1}R^{a\dots
c_{n-4},d}+\dots +\delta_b^d R^{c_1\dots c_{n-4},a}\, .
\end{equation} These also follow from the analysis of Section~3.  From  
these
commutators, one gets  the following relations
\begin{eqnarray} H_m&=&\delta_m^a (K^a{}_a-K^{a+1}{}_{a+1})\qquad
a=1,\dots,n-1\,
\nonumber\\  H_{n}&=&-(K^1{}_1 +K^2{}_2+K^{3}{}_{3}) + K^{n}{}_{n}\, .
\end{eqnarray}

The metric $G_{ij}$  has the universal form
\begin{equation} {\bf G^{+++}}={\bf I}_D -
\frac{1}{2}{\bf\Xi}_D~~{\rm and \ its\  inverse }~~ {\bf G^{-1}_{  
+++}}={\bf I}_D -
\frac{1}{D-2}{\bf\Xi}_D\, .
\end{equation}

The Weyl transformations corresponding to simple roots in the gravity  
line
induce an exchange of neighbouring components of the diagonal metric.   
The Weyl
transformation for the simple root  $\alpha_n$ induces the changes
\begin{eqnarray} K^{\prime a}{}_a&=& K^a{}_a\quad a=1,2,3\nonumber\\
K^{\prime a}{}_a&=& K^a{}_a-H_n\quad a=4,\ldots ,n-1\qquad\qquad  
K^{\prime
n}{}_n= K^n{}_n-2H_n\, .
\end{eqnarray} On   the $p^a$ the effect of the Weyl transformation  is  
given by
\begin{eqnarray} &&p^{\prime a}=p^a+(p^4+\ldots +p^{n-1})+2p^n\qquad
a=1,2,3
\nonumber\\ && p^{\prime a}=p^a\qquad a=4,\dots ,n-1\nonumber\\
&&p^{\prime n}=-p^n-(p^4+\ldots +p^{n-1})\, .
\end{eqnarray}

Finally, we point out that the computation of the Weyl reflections  
generated by
the simple roots as well as the outer automorphisms can be performed  
similarly
for all the other simple  $G^{+++}$. From the Dynkin diagram and from  
Table I, one
isolates the  $SL(D)$ gravity line and  completes the Cartan subalgebra  
by the
required number of dilatons. Weyl reflections corresponding to the  
simple roots
of the gravity line always interchange neighbouring diagonal elements  
of the
metric. One verifies that in the space of the fields
$p^i =\{p^a, \Phi^u\}$, the metric $G^{+++}_{ij}$ has always the   
universal form
Eq.(\ref{very}).

\setcounter{equation}{0}
\section{Dimensional reduction revisited}

In this section we shall show how  the properties of dimensional  
reduction of
`maximally oxidised' theories \cite{cremmerjlp99,lambertw01} can be
understood in terms of the general framework laid out in Section~2.   
We consider the 
theories of gravity in the highest available dimension $D$,  possibly  
coupled to
dilatons and to suitable $n$-forms with well-chosen coupling to the  
dilatons,
which exhibit, upon dimensional reduction to three space-time  
dimensions, a
symmetry under a simple  Lie group $\cal G$ 
in its maximally non-compact form. For any simple Lie group  
$\cal G$,
there exists such a maximally oxidised theory described by a Lagrangian
$\cal L_G$.

We first present a review, which follows \cite{lambertw01}, of the  
essential
features of dimensional reduction from $D$ to three space-time  
dimensions.
  We then show how, for a special
class of  subgroups ${\cal G}^{(k)}$ in $\G$,  group invariant metrics  
${\bf
G^{(k)}}$ emerge from dimensional reduction on a
$k$-torus, with $D \ge k \ge 1$.  These generalise to all Lie groups  
the metrics
obtained in Eq.(\ref{partmetric})  for the
$E$-sequence    of $E_8^{+++}$ subgroups.  The third and crucial  part  
of the
section is dedicated to the formulation of Weyl-preserving embedding
equations relative to ${\cal G}^{(k)} \subset \G$. These play an  
important role in
the  following sections.

Dimensional reduction to $D-k$ dimensions is equivalent to  
compactification on a
$k$-torus of vanishing radii, so as to eliminate the massive  
Kaluza-Klein
excitations. Note that this does not mean that the  metric components
$g_{\alpha\beta} \,  (\alpha,\beta= D,D-1,\dots, D-k+1)$ of the
$k$-torus are small but rather that the coordinates $y^\alpha$ of the  
torus metric
$ds^2=g_{\alpha\beta}dy^\alpha dy^\beta$ are periodic with vanishing  
small
periodicity
$\lambda$. The insensitivity to massive Kaluza-Klein modes implies that  
the
dimensional reduced theory has a symmetry under the  deformation group  
of the
$k$-torus, namely $GL(k)$.

The dimensional reduction  from the $k$-torus is performed step by step  
so as
to remain in the Einstein frame in the remaining non-compact dimensions  
with a
standard kinetic term for the resulting scalar.  When $k$ reaches  
$D-3$, the
$GL(D-3)$ symmetry group is, for the theories defined by $\cal L_G$,  
enhanced
to a simple non-compact Lie group $\cal G$.  All the original fields of  
the theory
are transmuted to scalars and these form a non-linear realisation of  
the group on
the coset space $\cal G/  H$ where $\cal H$  is the maximal compact  
subgroup of
$\cal G$. Examples of  Lagrangians
$\cal L_G$ are Eqs.(\ref {Mth}),  (\ref {Bth}),   (\ref {Hth}) and  
gravity itself, which
yield at
$D=3$ respectively  ${\cal G}= E_8$,  ${\cal G}= D_{D-2}$, ${\cal G}=  
B_{D-2}$ and
$A_{D-3}$.

The group $\cal G$ is generated in the following way. After $k$ steps  
one gets,  in
addition to the dilatons $\Phi^u , u=1,2,...,q$ originally present,
$k$ scalars $\phi^{\hat a}$ that parametrise the radii of the  
$k$-torus.  One also
obtains scalars arising from dimensional reduction of the 2-forms
$F_{\mu\nu}$ generated at each step from the Kaluza- Klein reduction of  
gravity.
More scalars similarly arise after $n-1$ steps from $n$-forms present  
in the
original Lagrangian and, after $D-n-1$ steps, one also gets scalars by  
dualising
the
$n$-forms.  When $k$ reaches $D-3$,  all fields can be dualised to  
scalars. The final
result can be  described by a Lagrangian
\begin{equation}
\label{reduce} {\cal L} =R -\frac{1}{2} \partial_\mu {\varphi}^{\hat
\imath}\delta_{\hat \imath \hat \jmath}\partial^\mu  {\varphi}^{\hat
\jmath} -{1\over2}
\sum_{\vec
\alpha}e^{\alpha_{\hat \imath}  \varphi^{\hat \imath}}
\partial_\mu\chi^{\vec\alpha}\partial^\mu\chi^{\vec\alpha}+\dots \, ,
\end{equation} where
\begin{equation}
\label{equal}
\varphi^{\hat \imath} = \{\phi^{\hat a},\Phi^u\}\, \quad \hat  
a=D,\dots, 4\quad
u=1,\dots, q\,  .
\end{equation}
  Here the $\vec\alpha$ are Euclidean vectors with $D-3+q$ components
$\alpha_{\hat
\imath}$ and they are used to label the   additional scalars
$\chi^{\vec\alpha}$.

The scalar fields in the Lagrangian Eq.(\ref{reduce}) form a non-linear  
realisation
of a maximally non-compact  Lie group $\cal G$ in the coset space $\cal  
G/H$
parametrised by
\begin{equation}
\label{defg} g =
\sum_{\vec\alpha }e^{-\chi^{\vec\alpha} E_{\vec\alpha}} e^{-\frac{1}{2}
\varphi^{\hat\imath}  H_{\hat \imath}}\, .
\end{equation} The vectors $\vec\alpha$ are the positive roots of $\cal  
G$ and
the Cartan generators $H_{\hat\imath}$ are normalised by
\begin{equation}
\label{norm} Tr(H_{\hat \imath}H_{\hat \jmath})=2\delta_{\hat \imath  
\hat
\jmath}\, .
\end{equation} We write the $H_{\hat \imath}$-basis as
\begin{equation}
\label{Hbasis} H_{\hat \imath} = \{H_{\hat a},R_{\hat u}\}\, \quad \hat  
a=D,\dots,
4\quad u=1,\dots ,q
\end{equation} where $R_{\hat u}$ are the dilaton generators in this  
new basis.

The group ${\cal G}$ is a simple Lie group. This  follows from the  
structure of the
simple roots. We now recall in some detail how these arise. First  
consider the
scalars obtained from gravity alone. At the first step of the  
dimensional
reduction,  after Weyl rescaling to the Einstein frame and after  
rescaling the
$\phi^D$ scalar to get the canonical normalisation of the  kinetic  
term, the
Einstein action becomes
\begin{eqnarray}
\label{einstein}
\int d^Dx \sqrt{-g^{(D)}}R^{(D)} &=& \int d^{D-1}x
\sqrt{-g^{(D-1)}} \left (R^{(D-1)}\right.\nonumber\\ &-&\frac{1}{2}
\left.\partial_\mu\phi^D\partial^\mu\phi^D -{1\over 4}e^{-2(D-2)
\beta_{D-1}\phi^D}F_{\mu\nu}F^{\mu\nu}\right) ,
\end{eqnarray} where
\begin{equation}
\label{scale}
\beta_{D}=\frac{1}{\sqrt{2(D-1)(D-2)}}\, .
\end{equation}  The further steps proceed similarly and   simple roots  
arise at
the first reduction following the appearance of a 2-form.  In a sense,  
this is the
`fastest' way to obtain a
$\chi^{\vec\alpha}$ scalar in the step by step dimensional reduction  
process.
Reducing  $D-3$ dimensions, one gets in this way  $D-4$
$\chi^{\vec \alpha}$ scalars and the corresponding $\vec\alpha$ are the  
simple
roots of
$SL(D-3)=A_{D-4}$.   The components of these
  simple roots  in the $(D-3+q)$ dimensional orthonormal Euclidean basis  
defined
by Eq.(\ref{equal}) are
\begin{equation}
\label{root}
\alpha_{(\kappa)\hat\imath}= (\underbrace {0,\dots,
2(D-\kappa-3)\beta_{D-\kappa-1},-2(D-\kappa-1)\beta_{D-\kappa} ,0,\dots,
0}_{k=D-3 \rm~ terms}\,;\underbrace{0,\dots, 0}_{q\rm ~terms}) \, ,
\end{equation}
  where $\kappa=1,2, ...,D-4$ and there are $\kappa-1+q$ zeros on the  
right.

The  Dynkin diagram of $SL(D-3)=A_{D-4}$ and  the $D-3$ Cartan  
generators
$H_{\hat a}$ parametrised by the
$\phi^{\hat a}$  extend  the  subgroup  $SL(D-3)$ to the full  
deformation group
$GL(D-3)$. Note that we have omitted from the above consideration the  
scalars
that arise by dualising the `graviphotons'
$F_{\mu\nu}$ present in three dimensions. The reason is that they do not
generate simple roots, except when the original Lagrangian $\cal L_{G}$  
contains
only gravity, in which case one dualised graviphoton does.

When suitable $n$-forms and $q$ dilatons are present in the  
uncompactified
theory,  the enhancement of symmetry from $GL(D-3)$ to $\cal G$ appears
through
$q+1$ simple roots attached to the Dynkin diagram of  $SL(D-3)$.  The  
Dynkin
diagram of $\cal G$ is the part of the diagram of $\cal G^{+++}$ in  
Fig.1 and Fig.2
which sits on the right of the dashed line cutting off the three first  
nodes. Its
$SL(D-3)$ subgroup is the corresponding part of the gravity line  
$SL(D)$ in
$\cal G^{+++}$. A deeper relation between these two algebras in the  
framework
of maximally oxidised theories will emerge later in this section.

Let us first illustrate how the general properties of dimensional  
reduction are
realised for  the 11-dimensional M-theory, Eq.(\ref{Mth}), and for the  
bosonic
case, Eq.(\ref{Bth}).
\begin{itemize}
\item[] {\it a)  M-theory}. The additional simple root is due to the  
`electric'
4-form that gets attached to the gravity line after 4 steps. Its  
components in
the orthonormal Euclidean  basis are
\begin{equation}
\label{Mel}
  \alpha_{(e)\hat\imath}=2(D-5) (\underbrace { 0, \dots ,\beta_{D-3},
\beta_{D-2},\beta_{D-1} }_{k=D-3 \rm~ terms})\, .
\end{equation} The gravity line has $D-4=7$ nodes and the group $GL(8)$  
is
enhanced to the simple group $E_8$.
\item[]  {\it b)  The bosonic theory}. The first additional simple root  
is due to the
`electric' 3-form. It gets attached to the gravity chain after three  
steps.  The
second one stems from its `magnetic' dual
$(D-3)$-form and joins  the gravity line in  three dimensions.  The  
electric and
magnetic simple roots  components in the orthonormal Euclidean basis   
are
\begin{eqnarray}\label{Bel}
\alpha_{(e)\hat\imath}&=&(\underbrace{ 0,
\dots,2(D-4)\beta_{D-2},2(D-4)\beta_{D-1}}_{k=D-3 \rm~ terms}\, ;-l)\,  
,\\
\label{Bm}
\alpha_{(m)\hat\imath}&=&(\underbrace{ 0,4\beta_4,
\dots,4\beta_{D-2},4\beta_{D-1}}_{k=D-3
\rm~ terms}\, ; l)\, .
\end{eqnarray}  The gravity line has $D-4$ nodes and the group  
$GL(D-3)$ is
enhanced to the simple group $D_{D-2}$.
\end{itemize}

We now make precise the relation of dimensional reduction with the  
general
construction of Section~2 and derive the group invariant metrics ${\bf  
G^{(k)}}$.
To this effect, we relate
  the fields  $\varphi^{\hat \imath} = \{\phi^{\hat a},\Phi^u\}$  
obtained in the
dimensional reduction to the fields $p^i=\{p^a, \Phi^u\}$ in  
$D$-dimensional
space-time. This amounts essentially to `undo' the reduction process  
and to
consider, in D dimensions,  the moduli of a diagonal $D-3$-torus
$p^\alpha\, ,g_{\alpha\alpha}=
\exp(2p^\alpha)\eta_{\alpha\alpha}
$, for $\alpha= D, D-1,\dots 4$,  and diagonal metric
fields in the non-compact dimensions  $p^\mu\, ,g_{\mu\mu}=
\exp(2p^\mu)\eta_{\mu\mu}
$,
$\mu=1,2,3$. Thus we decompose the $p^a$ in two sets, namely  
$p^a=\{p^\alpha,
p^\mu\} $.

We first consider the $p^\alpha$. When performing a dimensional  
reduction step
by step from $D$ to
$D-k$ dimensions, the fields $\phi^{\hat a}$   are  shifted from $ p^a$  
by the Weyl
rescalings to the Einstein frame  and by the rescaling  fixing their  
kinetic term.
Iterating the steps according to Eq.(\ref{einstein}), one gets
\begin{equation}
\label{sigma}
  k\hbox{ terms} \left\{\begin{array}{l} p^D=- (D-3)  \beta_{D-1}\phi^D\\
p^{D-1}=-(D-4) \beta_{D-2}\phi^{D-1} + \beta_{D-1} \phi^D\\  
p^{D-2}=-(D-5)
\beta_{D-3}\phi^{D-2}+ \beta_{D-2} \phi^{D-1}+
\beta_{D-1}
\phi ^D\\ \vdots\\
p^{D-k+1}=-(D-k-2) \beta_{D-k}\phi^{D-k+1} +\dots + \beta_{D-1}  
\phi^D\, .
\end{array}\right.
\end{equation} For $i=D-k+1,\dots, D+q$~, comparing  Eq.(\ref{defg})  
with
Eq.(\ref{HK}),
\begin{equation} p^i L_i\, (\equiv p^\alpha K^{ \alpha}{}_\alpha  
+\Phi^u R_u)=
-\frac{1}{2}\varphi^{\hat\imath} H_{\hat\imath}\,(\equiv -\frac{1}{2}
\phi^{\hat a} H_{\hat a} -\frac{1}{2}
\Phi^u R_{\hat u})
\, ,
\end{equation}  one has
\begin{equation}
\label{dilatonnorm} p^\alpha K^{ \alpha}{}_\alpha  
=-\frac{1}{2}\phi^{\hat a}
H_{\hat a}\quad,\quad R_u =  -\frac{1}{2}R_{\hat u}\, ,
\end{equation} and we may   express the $K^{ \alpha}{}_\alpha $ in  
terms of the  $
H_{\hat a}$ as
\begin{equation}
\label{KH} K^{ \alpha}{}_\alpha = H_{\hat a} a^{\hat a}_{~\alpha}\, .
\end{equation}
  The matrix ${\bf a}=a^{\hat a}_{~\alpha}$ is equal to
$ ({2\bf  b})^{-1}$ where ${\bf b}$ is the matrix relating the
$p^\alpha$ to the $\phi^{\hat a}$ in Eq.(\ref{sigma}). This yields the  
triangular
matrix
\begin{equation}
\label{matrix} {\bf a}= \left[\begin{array}{ccccccc}
\frac{1}{2(D-3)\beta_{D-1}}&0&0&.&.&.&0\\ \beta_{D-2}
&\frac{1}{2(D-4)\beta_{D-2}}&0&.&.&.&0\\
\beta_{D-3}&\beta_{D-3}&\frac{1}{2(D-5)\beta_{D-3}}&.&.&.&0\\
\beta_{D-4}&\beta_{D-4}&\beta_{D-4}&. &.&.&0\\.&.&.&.&  
.&.&0\\.&.&.&.&.&.&0\\
\beta_{D-k}&\beta_{D-k}&\beta_{D-k}&.&.&.&\frac{1}{2(D-2-k)\beta_{D- 
k}}\\
\end{array}\right]
\end{equation} For $k=D-3$, the   ${\cal G}$-invariant metric in the  
space of  the
$p^i$ fields can be obtained, as in Eq.(\ref{trace}), using
\begin{equation}
\label{compute}
  G^{D-3}_{ij} = {\rm Tr }(L_iL_j)= \frac{1}{4} {\rm Tr} R_{\hat  
u}R_{\hat
v}\delta^{\hat u}_{~u}\delta^{\hat v}_{~v}+ {\rm Tr }(H_{\hat a}H_{\hat  
b})a^{\hat
a}_{~\alpha}a^{\hat b}_{~\beta}\, .
\end{equation} The normalisation given in equation Eq.(\ref{norm})  
yields from
Eq.(\ref{matrix})
\begin{equation}
\label{modmetric3} {\bf G^{(D-3)}}=\frac{1}{2}{\bf I}_q \oplus( {\bf  
I}_{D-3} +
{\bf\Xi}_{D-3})~,~ {\bf G_{(D-3)}^{-1}}=2\, {\bf I}_q \oplus( {\bf  
I}_{D-3} -
\frac{1}{D-2} {\bf\Xi}_{D-3}) \, .
\end{equation} The metric Eq.(\ref{modmetric3})   can  be generalised  
for
$k$-torus compactifications with $k\ne D- 3$. For $k<D-3$, we shall  
only consider
the sequences of simple groups ${\cal G}^{(k)}=E_8^{(k)},
D_{D-2}^{(k)},B_{D-2}^{(k)}$ of rank $k$  obtained by deleting nodes on  
the gravity
line of ${\cal G}=E_8, D_{D-2},B_{D-2}$, together with the magnetic  
node attached
to it in the $D$ and $B$ series. Such attachment occurs at $k=
(D-3)$.\footnote{One could also consider non-simple groups by keeping  
for
instance the magnetic root for $k=D-4$. Indeed,  the corresponding  
enhancement
of  symmetry  appears in the
$B$ and $D$-series  by dualising a three form when
$k=D-4$, but the extra node gets attached to the gravity line only when  
$k$
reaches
$D-3$. For simplicity we shall restrict the detailed analysis  to  
simple groups
${\cal G}^{(k)}$, but the foregoing discussion can be easily extended  
to such
non-simple groups.} The metric for the $p^\alpha$ moduli of the
$k$-torus is obtained as above from Eq.(\ref{matrix}). One  
gets\footnote{Note that
${\bf G}$  is singular in the affine case $k=D-2$, but that its matrix  
elements
between roots Eq.(\ref{rootG}) are well defined provided they are  
computed by
regularising $D$  as $D+\epsilon$ and taking the limit
$\epsilon \to 0$ at the end. The  inverse metrics
${\bf G}^{-1}=(G^{-1})^{ij}$ are well defined for all $k>2$ and no  
regularisation is
needed to compute scalar products in the affine case. The matrix
${\bf G}^{-1}$ has, as required by the theory of Kac-Moody algebras,  
positive
eigenvalues for
$k\le (D-3)$, one zero
  eigenvalue for $k=D-2$, and one negative
  eigenvalue for $k=D-1$ and $k=D$.}
\begin{equation}
\label{modmetricp} {\bf G^{(k)}}=\frac{1}{2}{\bf I}_q \oplus( {\bf I}_k
-\frac{1}{k+2-D} {\bf\Xi}_k)~,~ {\bf G_{(k)}^{-1}}=2\, {\bf I}_q  
\oplus( {\bf I}_k -
\frac{1}{D-2} {\bf\Xi}_k) \, .
\end{equation}
  It is important to realise that one may  obtain the same equation from
  the general method of Section~2.  One embeds   ${\cal
G}^{(k)}$ into ${\G}$ by deleting the nodes  $m$  in the Dynkin
diagram of  ${\G}$ which are not in the Dynkin diagram of  ${\cal
G}^{(k)}$. In the general formalism of Section~2, this amounts to
equate to zero  the field
$q^m$ parametrising the Cartan generator
$H_m$ in the Chevalley basis, as in the derivation of the
metric Eq.(\ref{partmetric}) for the $E$ sequence.  This method  
validates the
extension of Eq.(\ref{modmetricp}) for the case where
$D\ge k > D-3$.
However, it is  interesting to obtain this latter extension from
dimensional reduction directly. This
exercise exhibits the universality of the metrics ${\bf
G^{(k)}}$  up to ${\bf
G^{+++}}$.

We cannot perform  dimensional
reduction on a
$k$-torus for $k>(D-3)$ in a straightforward way.  Gravity does not  
exist below
three space-time dimensions and no new independent scalar could emerge  
by
further compactifications. However  the rank of the torus deformation  
group
$GL(k)$  increases  up to $k= (D-1)$ when all space dimensions are  
compactified,
and  to $k=D$ if  time is compactified as well.  These deformations add  
new nodes
to the gravity line  obtained from dimensional reduction to three  
dimensions.
The resulting Dynkin diagrams  define, for
$k=(D-2), (D-1)$ and $D$,  the affine Kac-Moody algebra
${\cal G}^+ $, the Lorentzian overextended
  ${\cal G}^{++}$ and the very extended ${\cal G}^{+++}$ Kac-Moody  
algebras.  We
shall see that all the simple roots of all such Kac-Moody algebras in  
the
$L$-basis  Eq.(\ref{Kbasis}) are determined by a dimensional  
regularisation
procedure even below three dimensions, and so are all the corresponding
metrics Eq.(\ref{modmetric}).

We note that, for
$k=D-3$, we may compute the roots of the gravity line in the $L$-basis   
  from
Eqs.(\ref{KH}) and (\ref{root}). One gets, for a given root, two  
non-vanishing
commutators
\begin{eqnarray}
\nonumber [K^\alpha{}_\alpha,E_{\vec \alpha}]&=&a^{\hat a}_{~\alpha}[  
H_{\hat
a},E_{\vec\alpha} ]=+ E_{\vec\alpha}\, ,\\
\label{second} [K^{\alpha+1}_{~\alpha+1},E_{\vec\alpha}]&=&a^{\hat
a}_{~a-\alpha+1}[ H_{\hat a},E_{\vec\alpha} ]=- E_{\vec\alpha}\, .
\end {eqnarray} These are indeed the components of the roots defined by  
the
commutation relations of the operators $K^\alpha_{~\beta}$ given in
Eq.(\ref{Kcom}). It is easily seen that we still obtain the correct  
commutation
relations for these operators by {\em defining} $\phi^{\hat a}$ for  
${\hat a }<3$
through Eq.(\ref{sigma}) and regularizing
$\beta_2,\beta_1$ and $\beta_0$.  Namely one replaces $D$ by
$D+\epsilon$ and takes the limit $\epsilon\to 0$ at the end.  All the  
other simple
roots can  be computed as above in the
$L$-basis at  $k=D-3$, using for their components in the dilaton space  
the second
equation in Eq.(\ref{dilatonnorm}). These roots  are not affected by the
extensions. For instance the root components in the
$H_{\hat
\imath}$-basis obtained from Eqs.(\ref{Mel}), (\ref{Bel}) and  
(\ref{Bm}) are in the
$L$-basis, for
$E_8^{+++}$   and for
$D_{D-2}^{~+++}$,
\begin{equation}
\label{KMel}
  \alpha_{(e) i}= (\underbrace {0,\dots,0, 1, 1,1, }_{k=D \rm~ terms})\,  
,
\end{equation} and
\begin{eqnarray}\label{KBel}
\alpha_{(e)i}&=&(\underbrace{0, \dots ,0,1, 1}_{k=D\rm~ terms}\,  
;\frac{1}{2} l)\,
,\\\label{KBm}
\alpha_{(m)i}&=&(\underbrace{0,0,0,0,1, 1,\dots,1}_{k=D
\rm~ terms}\, ; -\frac{1}{2} l)\, .
\end{eqnarray} The relations Eqs.(\ref{second})-(\ref{KBm}) are in  
agreement
with the commutation relations in the $L$-basis listed in the examples  
of
Section~2.

Thus the relations Eqs.(\ref{KH}) with regularised coefficients $a^{\hat
a}_{~\alpha}$ remain valid when
$k>D-3$.   In addition, extending   the normalisation Eq.(\ref {norm})  
to
$k>D-3$,
  one verifies that
  the regularised  scalar products of all simple roots are well defined.  
What
happens is that the  regularised roots  acquire one imaginary  
components when
$k$ reaches $D-1$, thereby switching from the Euclidean scalar product
$\delta_{\hat
\imath \hat
\jmath}$ in Eq.(\ref{norm}) to a Lorentzian one $\eta_{\hat
\imath \hat
\jmath}$ for  ${\cal G}^{++}$ and ${\cal G}^{+++}$. The computation
Eq.(\ref{compute}) thus remains  valid and the regularised metric
Eq.(\ref{modmetricp}) is recovered  in this way for
$k>D-3$.

We  now  formulate the Weyl-preserving embedding equations that
play a key role in unveiling the new symmetries of Section~4, and the  
dualities of
Section~5. So far we have only considered the $k$ fields  $p^\alpha,  
~\alpha=
D-k+1,\dots, D$ but we shall now  relate them to the fields   
$p^\mu,~\mu= 1,\dots,
D-k$ in  non-compact dimensions. As pointed out above,
the embedding of  ${\cal G}^{(k)}$ into ${\G}$ may be viewed as the
result of deleting nodes $m$ in the Dynkin diagram
of $\G$.  This means that the fields $q^m$ parametrising the Cartan
generator $H_m$ in the Chevalley basis are equated to zero. Such
embedding relates the $D-k$ fields
$p^\mu$  in the non-compact dimensions to the
moduli $p^\alpha$  of the compactification torus,  and  possibly to the
dilatons. The relation between fields resulting from the embedding of a
subgroup  ${\cal G}^{(k)}$  obtained by putting Chevalley fields $q^m$  
to zero are
general and  not a specific feature of the dimensional reduction of an  
oxidised
theory. In view of their relevance in Sections~4 and 5 we  shall thus  
use the
generic labelling of the physical fields $p^i=\{p^a,\Phi^u\}$, and the  
label $(k)$ in
${\cal G}^{(k)}$ will refer in general to the subgroup obtained by  
deleting the first
$D-k$ nodes of the gravity line together with any magnetic node possibly
attached to it.

The  fields $p^i$ are given in terms of  the fields
$q^m$ by the relation Eq.(\ref{HK}), namely $p^a  =  q^m r_m^{~a}$,
$\Phi^u= q^m r_m^{~u}$. Putting $q^m=0$ for $m=1,\dots, D-k$ and  
possibly
$q^x=0$ where
$x$ labels a magnetic node,  one obtains
relations between the fields
$p^i=\{p^a,\Phi^u\}$   defining the embedding of ${\cal G}^{(k)}$ into
${\cal G}^{(+++)}$. To derive these relations we shall use the equality  
between
the quadratic forms in the $q^m$ and the $p^i$ variables, namely
\begin{equation}
\label{mapping} p^iG^{(k)}_{ij}p^j=  
p^aG^{(k)}_{ab}p^b+\frac{1}{2}\sum_{u=1}^q
(\Phi^u)^2=q^m \, G_c^{(k)}{}_{mn}\, q^n
\end{equation} where the matrix $ G_c^{(k)}{}_{mn}$ for  ${\cal  
G}^{(k)}$ is given in
Eq.(\ref{groupmetric}). As seen from Eq.(\ref{modmetric}), this  
equality appears
valid only if the matrix ${\bf r}$ is invertible, that is if the rank  
of $G^{(k)}_{ij}$
and $ G_c^{(k)}{}_{mn}$ are equal. This condition is clearly satisfied   
if the only
deleted nodes belong to the gravity line as each deleted node reduces  
the rank
of both matrices by one unit. It is
   also satisfied when one deletes the magnetic node attached to the  
gravity line
together with the gravity node to which it is glued.  The first case  
occurs for
example in the embedding of the $E^{(k)} $ sequence considered in   
Section~2.2 or
for any embedding of  ${\cal G}^{(k)}$ in
${\G}$ when
$k=D-1$,
$D-2$ and  $D-3$. As mentioned above the second case occurs in the $D$  
and
$B$ series when
$k < D-3$, that is when the  magnetic node glued to the fourth root of  
the gravity
line is deleted.  The corresponding equation, $q^{D+1}=0$, yields a  
relation
between the dilaton and the $p^a$ fields which ensures that  
Eq.(\ref{mapping})
relates expressions with the same number of independent variables.

Consider the embedding ${\cal G}^{(k)}\subset {\cal G}^{+++}$ and the  
sequence
$q^m=0$ for $m=1,\dots, D-k$ of deleted nodes on the gravity line. It  
follows
from Eq.(\ref{mapping}) that we get $D-k$ relations
\begin{equation}
\label{pmap} p^iG^{(s+1)}_{ij}p^j=p^kG^{(s)}_{kl}p^l \, ,
\end{equation} for $s = k,k+1,\dots, D-1$, and one additional relation  
between
the dilaton  and the $p^a
$ fields if a magnetic node attached to the gravity line is deleted.   
From
Eq.(\ref{modmetricp}) we get the explicit relation
\begin{eqnarray}
\label{mapp}  (p^{D-s})^2 +  
\sum_{a=D-s+1}^D(p^a)^2&-&\frac{1}{s+3-D}~(p^{D-s}
+
\sum_{a=D-s+1 }^D p^a)^2\nonumber\\
=\sum_{a=D-s+1}^D(p^a)^2&-&\frac{1}{s+2-D}~(
\sum_{a=D-s+1}^D p^a)^2\, .
\end{eqnarray} Note that it follows from the relation Eq.(\ref {HK})  
between the
$p^i$ and the $q^m$ fields, and from footnote 5, that Eq.(\ref{mapp}) is
well-defined (after regularisation) even when  $s= D-2$ or $D-3$. One  
gets from
Eq.(\ref{mapp})
\begin{equation}
\left[ p^{D-s} -\frac{
1}{s+2-D}\sum_{a=D-s+1}^Dp^a\right]^2\frac{s+2-D}{s+3-D}=0\, .
\end{equation} To obtain the embedding of  ${\cal G}^{(k)}$ into
  ${\cal G}^{+++}$, it suffices to use the  sequence of embedding  
relations
\begin{eqnarray}
\label{transform} {\cal G}^{++}\subset {\cal G}^{+++}~{\rm or}~
(s=D-1):\quad\qquad
\qquad\qquad\quad
\qquad \qquad  p^1&=&\sum_{a>1} p^a\qquad~~\nonumber\\ {\cal  
G}^{+}\subset
{\cal G}^{++}~{\rm or}~ (s=D-2):~ \qquad\qquad\quad 0=\sum_{a>2}  
p^a\quad
\rightarrow \quad  p^1&=&p^2\nonumber\\ {\cal G}^{D-3}\subset {\cal  
G}^{+}:\
\,\quad\qquad \qquad \qquad\qquad~\, -p^3=\sum_{a>3} p^a\quad
\rightarrow\quad   p^1&=&p^2\nonumber\\ {\cal G}^{D-4}\subset {\cal  
G}^{D-3}:
~~~~\,  \qquad
\qquad\qquad\quad  -2 p^4=\sum_{a>4} p^a\quad \rightarrow\quad
p^3&=&p^4\nonumber\\ &\vdots&\nonumber\\ {\cal G}^{(k)}\subset {\cal
G}^{(k+1)}:  \ \, \,  -(D-2-k)p^{D-k}=\sum_{a>D-k} p^a~ \rightarrow\quad
p^{D-k-1}&=&p^{D-k}\, ,
\end{eqnarray} and possibly the additional relation between the dilaton  
and the
$p^a$ encoded in the relation $\Phi^u= q^m r_m^{~u}$. A direct computing of the
embedding relations for the $p^a$ using Eq.(\ref{HK}) yields the relation $p^2=p^3$ for 
${\cal G}^{D-3}\subset {\cal G}^{+}$.

The metrics $G^{(k)}_{ij}$ are invariant under the generators of the  
Weyl group
and under the outer automorphisms  of the algebra ${\cal G}^{(k)}$. The
embedding Eq.(\ref{transform}) guarantees that these transformations are
uplifted to the  Weyl generators and outer automorphisms of
$\G $ associated with the  roots common to the two algebras. The reader  
may
verify that  the embedding equations do indeed determine the Weyl
transformations of $\G $  from those of ${\cal G}^{(k)}$ in all the  
examples listed
in Section~2.

We list particular solutions of the  embedding equations for the  
physical fields.
Let us first consider the case, studied in the present section, where  
${\cal
G}^{(k)}$ arises from a compactification on a
$k$-torus. A particular solution of Eq.(\ref{transform}) is obtained by  
taking all
$p^\mu$ equal to a constant $C$. We then get
\begin{equation}
\label{compact} C= {1\over k+2-D} \sum_{\alpha=D- k+1}^D p_\alpha\qquad  
k\ne
D-2\, .
\end{equation}
  This relation will be used in Section~5. Another useful relation is  
the one
characterised by the embedding  ${\cal G}^{++}\subset {\cal G}^{+++}$  
which is
just the first equation in  Eq.(\ref{transform}). For  convenience we  
list it
separately here
\begin{equation}
\label{embed1} p^1=\sum_{a=2}^D p^a\, .
\end{equation} This relation will be used in Section ~4.

\setcounter{equation}{0}
\section{  ${\cal G}^{+++}$ from Kasner-type solutions }

Given a theory  invariant under an algebra,  it must be the case that
any  solution of the equations of motion is  transformed into another
solution under the action of the algebra.  Conversely,  by carrying out  
symmetry
transformations on a known solution, and checking whether or not the
transformed expression still satisfies the equations of motion of the  
theory, one
provides evidence for an underlying symmetry, or concludes the theory  
does not
possess that symmetry.  As  pointed out before it
has been conjectured  that $ E_8^{+++}$ is a symmetry of M-theory and  
that
$D_{24}^{+++}$ is a symmetry of the bosonic string \cite{west01}. In  
this section
we will consider  particular classes of solutions not only of the  
effective actions
of these theories, but of all maximally oxidised theories $\cal L_G$  
described in
Section~3.   We will show that given any  such solution in  $\cal L_G$,  
the Weyl
group of
$\G$ transforms it  to another one and that  these solutions form a
representation of the group of Weyl transformations and of  outer
automorphisms of $\G$. This provides evidence for these discrete  
symmetries
and indicates the relevance of the $\G$ algebra for  the theory defined  
by $\cal
L_G$ .

All these field theories contain gravity and also in  some cases
  a number of scalar fields $\Phi^u$ as well as other fields. We will
  consider solutions for which only the diagonal components of the  
metric and the
scalar fields are non-zero. As  explained in Sections~2 and 3 one may
associate to each such theory  a very extended algebras  ${\cal  
G}^{+++}$, and
taking only these fields to be non-trivial corresponds to selecting  a
Cartan subalgebra. We have  calculated the action of the Weyl group of
${\cal G}^{+++}$ on these  fields in Section~2. As explained there,  
although  one is
considering only gravity and scalars fields,  one is probing the full  
Weyl group of
${\cal G}^{+++}$.

The non-trivial sector of the theory has an action of the form
\begin{equation}
\int d^D x \sqrt{-g}\left(g^{\mu\nu}R_{\mu\nu}-
\frac{1}{2}\sum_{u=1}^q   
g^{\mu\nu}\partial_\mu\Phi^u\partial_\nu\Phi^u\right)\, .
\end{equation} Varying the metric as well as the dilatons, we find the  
equations
of motion can be written as
\begin{equation} R_{\mu\nu}=\frac{1}{2}\sum_{u=1}^q \partial_\mu \Phi^u
\partial_\nu \Phi^u
\qquad,\qquad \partial_\mu(\sqrt{-g}g^{\mu\nu}\partial_\nu\Phi^u)=0  
\quad
u=1,...,q\, .
\end{equation} Putting the off-diagonal components of the metric  to   
zero,
   we write, following Section~2 and 3,  its diagonal components  as
$g_{aa}=\eta_{aa} e^{2p^a},\  a=1,\ldots ,D$. It is straightforward to  
calculate the
form of the above equation of motion for this diagonal metric. One  
finds that
\begin{eqnarray}
\label{curvature} R_{ab}&=&\{-\partial_a\partial_b\sum_c
p^c+\partial_a\partial_b(p^a+p^b)-\sum_c \partial_a p^c\partial_b
p^c\nonumber \\&&+\sum_c \partial_a p^b\partial_b p^c +\sum_c
\partial_a p^c\partial_b p^a -2 \partial_b p^a\partial_a
p^b\}e^{-(p^a+p^b)}\nonumber\\ &&-\eta_{ab}\sum_c \eta^{cc}
\{\partial_c\partial_c p^a+
\partial_c p^a \partial_c \sum_d p^d -2 \partial_c p^a \partial_c
p^c\}e^{-2p^c}\nonumber\\ &=&\frac{1}{2}  e^{-(p^a+p^b)}\sum_{u=1}^q   
\partial
_a
\Phi^u  \partial_b \Phi^u\, ,
\end{eqnarray}  and
\begin{equation}
\sum_{a,b}\eta^{ab}\partial  
_a(\exp[\sum_cp^c-p^a-p^b]\partial_b\Phi^u)=0\, .
\end{equation}

Let us  denote  the coordinates of our manifold by $(v,x^a),\  
a=2,\ldots D$ and
adopt for $p^a$  the simple dependence
$p^a=\tilde p^a v$.  This yields a  line element and dilatons of the  
form
\begin{eqnarray}
\label{ukasner} ds^2&=& -e^{2\tilde p^1  v}d (\tilde p^1v)^2 +
\sum_{a=2}^D e^{2\tilde p^a v}~(dx^a)^2\, ,
\\
\label{udilat}\Phi^u &=& {\tilde p^u}\, v\quad u=1,\dots q\, .
\end{eqnarray} We find this is a solution of the above equations of  
motion if and
only if
\begin{equation}
\label{uconstraint}
\tilde p^1=\sum_{a=2}^D\tilde     p^a	\, ,
\end{equation}
\begin{equation}
\label{uequation}
\sum_{a=2}^D(\tilde p^a)^2  - \left(\sum_{a=2}^D\tilde p^a\right)^2
+\frac{1}{2}\sum_{u=1}^q (\tilde p^u)^2=0 \, .
\end{equation} The first of these constraints arises from the space  
components
$R_{aa}$ equations while the latter arises from the time component  
$R_{vv}$. The
equation of motion for $\Phi^u$ leads to no new constraints.

According to the analysis of Section~2,  the group parameters
$\{\tilde p^1v,\tilde p^a v; \tilde p^u v\}$ entering the Kasner  
solutions  given by
Eqs.(\ref{uconstraint}) and (\ref{uequation})  can be viewed as fields
parametrising  the abelian   subalgebra of
$ {\cal G}^{+++}$ in the
$L$-basis. Here $ {\cal G}^{+++}$ is the very extended Kac-Moody  
algebra of any
Lie group defined in $D$ space-time dimensions and  involving $q$  
dilatons
according to Table I.  The subset of fields $p^i=\{\tilde p^a v;\tilde  
p^u v\}$,
$(a=2,\dots D;u=1,\dots q)$, are the independent moduli  of the Kasner  
solutions.
Comparing the Einstein equation Eq.(\ref{uequation})
  with Eq.(\ref{modmetricp}), we see that  the  moduli $p^i$   satisfy
\begin{equation}
\label{invover}
  p^iG^{++}_{ij}p^j=0\, .
\end{equation}
  In this equation we recognise the group invariant metric ${\bf  
G}^{++}$ of the
overextended algebra  $ {\cal G}^{++}$ restricted to its Cartan  
subalgebra This
metric  is invariant under the group $S({\cal G}^{++})$ generated  by  
the Weyl
reflections $W_\alpha ({\cal G}^{++})$ associated with  the simple  
roots  $\alpha$
of $ {\cal G}^{++}$ and by the
   outer automorphisms of the Dynkin diagram of $ {\cal G}^{++}$. It  
then follows
from Eq.(\ref{invover}) that  the moduli space
  of the Kasner solutions can be decomposed into linear representations  
of
$S({\cal G}^{++})$.

On the other hand, {\em the Einstein constraint Eq.(\ref{uconstraint})  
is identical
to the embedding equation Eq.(\ref{embed1})}, and  we learn from the  
general
analysis of embeddings given in Section~3 that $ {\cal G}^{++}$ sits in
$\G$ as the regular embedding  obtained by deleting its first gravity  
node
$\alpha_1$. As discussed there, the generators of  $S({\cal G}^{++})$
  are identified by this embedding to the corresponding generators of   
$S({\cal
G}^{+++})$, because they leave  both metrics ${\bf G}^{++}$ and ${\bf
G}^{+++}$ invariant. One may verify that Eq.(\ref{uconstraint}) and
Eq.(\ref{invover}) imply that the moduli $\hat p^k=\{\tilde p^1v,\tilde  
p^a
v;\tilde p^u v\} $ satisfy
\begin{eqnarray}
\label{invery} \hat p^k G^{+++}_{k
l}\hat p^l&=&\sum_{a=2}^D(\tilde p^a v)^2 + (\tilde p^1v)^2  -
\frac{1}{2} \left(\sum_{a=2}^D\tilde p^a v+ \tilde p^1v\right)^2
+\frac{1}{2}\sum_{u=1}^q (\tilde p^u v)^2\, ,\nonumber\\ &=&
p^iG^{++}_{ij}p^j=0\, .\hskip 6,65cm
\end{eqnarray} Each element of $S({\cal G}^{++})$ acting on the fields
$p^i$ is uplifted to an element of $S({\cal G}^{+++})$ through the  
embedding
Eq.(\ref{uconstraint}).  Note that
  the additional modulus $\tilde p^1 v$ is  preserved under all the Weyl
reflections which simply interchange two $p^i$ in the gravity line of $  
  {\cal
G}^{++}$. The  Weyl generators in $ {\cal G}^{++}$, associated to   
electric and
magnetic roots, as well as the outer automorphisms, act less trivially  
and do
change the modulus $\tilde p^1v$. Nevertheless they are also uplifted  
to Weyl
generators and automorphisms  of $ {\cal G}^{+++}$, as we  illustrate  
now for
M-theory, Eq.(\ref{Mth}), and the bosonic  theory, Eq.(\ref{Bth}).  A  
similar analysis
can be carried out for all very extended Kac-Moody algebras ${\cal  
G}^{+++}$.

\begin{itemize}
\item[]{\it a)  M-theory}.  There is no dilaton and the only  
non-trivial Weyl
reflection generator  in
${E_8}^{++}$  is
$W_{\alpha_{11}}$ where
$\alpha_{11}$ is the electric root. It is given by  Eqs.  
(\ref{WeylMe2}) and
  (\ref{WeylMe1}) for  $a\ne 1$.  The Weyl transform
$p^{\prime 1}$ of $p^1$ is given in terms of the Weyl transforms  
$p^{\prime a}$ of
$p^a$ by Eq.(\ref{uconstraint}), and we thus get, using   Eqs.  
(\ref{WeylMe2}) and
  (\ref{WeylMe1}),
\begin{equation}
  \tilde p^{\prime 1} =\sum_{a=2}^{11}\tilde p^{\prime a}= \tilde p^1 +
\frac{1}{3}(\tilde p^{9}+\tilde p^{10}+\tilde p^{11})\, .
\end{equation} As seen from  Eq.(\ref{WeylMe1}), this is indeed the  
correct
transformation of the uplifted Weyl reflection $W_{\alpha_{11}}$ in  
$E_8^{+++}$, in
accordance with the preceding discussion.

\item[]  {\it b)  The bosonic theory.} There is one dilaton and  the  
non-trivial
generators are the Weyl reflections in $D_{D-2}^{++}$ associated with  
the
electric root
$\alpha_{n-1}$, the magnetic root $\alpha_n$ and the outer isomorphism
exchanging the  roots
$\alpha_{n-1}$ and
$\alpha_{n-2}$. We examine each of these transformations.

  For $W_{\alpha_{n-1}}$
  the Weyl transforms of $\tilde p^a v$ and $\Phi=\tilde pv$ are given by
Eqs.(\ref{Del2})-(\ref{Del0}) for $a\ne 1$. Applying   
Eq.(\ref{uconstraint}) one
finds
\begin{equation}
  \tilde p^{\prime 1} =\sum_{a=2}^{n-1} p^{\prime a}=  \tilde  
p^1+{2\over (D-2)}(
\tilde p^{n-2}+ \tilde p^{n-1}) +{l\over D-2} \tilde p
\, ,
\end{equation} which, according to Eq.(\ref{Del2}) is,  as expected,   
the correct
transformation of the uplifted Weyl reflection in  $D_{D-2}^{+++}$.

Similarly, for the Weyl generator $W_{\alpha_n}$ corresponding to the  
magnetic
root $\alpha_n$, the Weyl transforms of $\tilde p^a v$ and $\Phi$ are  
read off
Eqs.(\ref{Dm2})-(\ref{Dm0}) for $a\ne 1$, and we now get from
Eq.(\ref{uconstraint})
\begin{equation}
\tilde p^{\prime 1} =\sum_{a=2}^{n-1} \tilde p^{\prime a}= \tilde  
p^1+{D-4\over
D-2}(
\tilde p^{5}+\dots + \tilde p^{n-1}) -{l\, (D-4)\over 2\, (D-2)} \tilde  
p,
\end{equation} which, as seen from Eq.(\ref{Dm2}),  is again the correct
transformation of the uplifted Weyl reflection in  $D_{D-2}^{+++}$.

Finally  the transformations of $\tilde p^a v$ and $\Phi$ under the  
outer
automorphism are given in Eqs.(\ref{Dout2})-(\ref{Dout0}) ($a\ne 1$)  
and hence,
from  Eq.(\ref{uconstraint}),
\begin{equation}
\tilde p^{\prime 1} =\sum_{a=2}^{n-1} \tilde p^{\prime a} = \tilde p^1  
+{2\over
(D-2)}
\tilde p^{n-1}+{l\over 2\, (D-2)} \tilde p \, ,
\end{equation} which agrees indeed with the outer automorphism in
$D_{D-2}^{+++}$.
\end{itemize}

To summarise, the moduli  $p^i$ of the Kasner solutions form linear
representations of the  group $S({\cal G}^{++})$ which is the subgroup  
of
$S({\cal G}^{+++})$ generated by all its outer automorphisms and by all  
its Weyl
generators except $W_{\alpha_1}$ associated with the root
$\alpha_1$.  This property  suggests the construction of  an enlarged  
set of
Kasner-like solutions whose moduli  do fall in  representations of
$S({\cal G}^{+++})$.
  We construct such an enlarged set by considering the following metrics
\cite{booklifschitz} which are
  analogous to the Kasner metrics but with the role of time and $x^2$
interchanged. We write
\begin{eqnarray}
\label{vkasner} ds^2&=& -e^{2\tilde p^2 v}(dx^2)^2 + e^{2 \tilde p^1  
v}d (\tilde
p^1 v)^2 +\sum_{a=3}^De^{2\tilde p^a v}~(dx^a)^2\,,
\\
\label{vdilat}\Phi^u &=& {\tilde p^u}\, v\qquad u=1,\dots q
\end{eqnarray} where $x^2$ is now a time variable and $v$ a space-like  
one.
Eqs.(\ref{vkasner}) and (\ref{vdilat}) solve Einstein equations if and  
only if,
\begin{equation}
\label{vconstraint}
\tilde p^2=\tilde p^1-\sum_{a=3}^D\tilde     p^a	\, ,
\end{equation}
\begin{equation}
\label{vequation} (\tilde p^2)^2+\sum_{a=3}^D(\tilde p^a)^2  -  
\left(\tilde
p^2+\sum_{a=3}^D\tilde p^a\right)^2 +\frac{1}{2}\sum_{u=1}^q (\tilde  
p^u)^2=0
\, .
\end{equation} Similar solutions can be written by singling out any
$x^a$ instead of $x^2$. Eqs.(\ref{vconstraint}) and (\ref{vequation})  
are identical
to the  constraints   Eqs.(\ref{uconstraint}) and (\ref{uequation}) for  
the Kasner
solutions, although Eq.(\ref{vconstraint}) can now be obtained from the  
time
component $R_{22}$ (or from any space component $R_{aa}, a>2$) and
Eq.(\ref{vequation}) from the particular space component $R_{vv}$  in
Eq.(\ref{curvature}). As previously the  equation for the dilaton does  
not
introduce new conditions.

  It follows  from (\ref{vequation}) that the moduli of the new  
solutions fall into
representations of a group
$\widetilde S({\cal G}^{++})$ isomorphic to $S({\cal G}^{++})$. The  
embedding in
$S(\G)$ defined by Eq.(\ref{vconstraint})  is however different. The  
moduli
$(\tilde p^2v,\tilde p^1v,\tilde p^3v,\dots,\tilde p^Dv;\tilde p^uv)$  
of the
solutions  Eqs.(\ref{vkasner}) and (\ref{vdilat})  differ from those of  
the Kasner
solutions Eqs.(\ref{ukasner})  and (\ref{udilat}) by the interchange of
$\tilde p^1$ with the new time modulus  $\tilde p^2$. This interchange  
can be
obtained from  applying the Weyl generator   $W_{\alpha_1}$ associated  
with the
root
$\alpha_1$ of $\G$.  $S({\cal G}^{++})$ and $\widetilde S({\cal  
G}^{++})$ are
conjugate in $S({\cal G}^{+++})$ by $W_{\alpha_1}$.  To see this, we  
first switch
$p^2$ and $p^1$, and then perform a general transformation of $S({\cal  
G}^{++})$.
The resulting transformation belongs to
$\widetilde S({\cal G}^{++})$ after relabelling $p^1$   as $p^2$. Thus  
we have
\begin{equation}
\label{conjugation}
\widetilde S({\cal G}^{++})= W_{\alpha_1}^{-1}S({\cal G}^{++})  
W_{\alpha_1}\, .
\end{equation} {\em We stress that $W_{\alpha_1}$ is the only generator  
of
$S({\cal G}^{+++})$ which was not a symmetry of the moduli space of the  
Kasner
solutions. We thus conclude that the enlarged set of Kasner-like  
solutions
constructed in this section span representations of $S({\cal  
G}^{+++})$}.

It is interesting to note that the cosmological billiards
\cite{damourhn00,damourh00,damourhjn01}, which arise as  time-dependent
solutions of Einstein's equations in the vicinity of a cosmological  
singularity for
the Lagrangians $\cal L_G$ considered in this paper, induce transitions  
between
Kasner solutions spanning linear representations of $S({\cal G}^{++})$.  
This
translates the fact that the billiards walls are invariant under
$S({\cal G}^{++})$, a feature which corroborates the possible existence  
of such a
symmetry at a fundamental level.  In addition the existence of an  
embedding
which uplifts all the elements of $S({\cal G}^{++})$ to elements of
$S({\cal G}^{+++})$ is encoded in Einstein theory, as discussed above.   
This fact is
not immediately apparent in the Hamiltonian formalism because the  
non-trivial
Weyl reflections, illustrated in the examples above,  affect the time  
modulus
$\tilde p^1 v$. As such they imply a redefinition of time which does  
not affect the
Hamiltonian constraint. The embedding is the crucial input which  
allowed us
  to build a representation of the full $S({\cal G}^{+++})$ from the  
enlarged set of
Kasner-like solutions, providing  evidence for a larger symmetry in all
oxidised theories.

\setcounter{equation}{0}
\section{String dualities}

It is well-known that  symmetries of  low energy effective actions of   
string
theories describe  symmetries of the perturbative string theories   and  
  appear
consistent with  their symmetries at the non-perturbative level. For  
instance
the effective action of the bosonic string, compactified on a $k$-torus  
has a
symmetry $O(k,k;R)$ realised non-linearly on the coset  
$O(k,k;R)/O(k)\times
O(k)$ which, in string theory, describes the moduli-space of the  
compactified
string.  Such distinct  vacua in perturbative string theory should  be  
related to
each other if a  background-independent non-perturbative formulation  
were
available. In addition the subgroup $O(k,k;Z)$ describes the   
perturbative
symmetry group of generalised    $T$-dualities (for a general review see
\cite{giveonpr}). The present approach, in which
${\cal G}^{+++}$ is viewed as a hidden symmetry of the  theory  $\cal
L_G$,  suggests that the discrete subgroup $S({\cal G}^{(k)})$ describes
  string  dualities in  cases where  $\cal L_G$ is the effective action  
of a
  string theory (see also references
\cite{elitzurgkr97,obersp98, banksfm98}), and that
${\cal G}^{(k)}$ is embedded in a larger symmetry group
$\G$.  We shall verify this connection between  $S({\cal G}^{(k)})$ and  
the duality
group of string theories, and we shall also check that the embedding of  
${\cal
G}^{(k)}\subset \G$, is precisely the one obtained in Section~3 by  
deleting nodes,
in accordance with the general construction of Section~2.

We shall analyse $M$-theory from the perspective of an
  $E_8^{+++}$ symmetry,  and  the bosonic string theory from that of a
$D_{24}^{+++}$ symmetry.  More generally we shall define  `string  
dualities' for
the whole
$D_{D-2}^{+++}$ series and compare them with those of the
$B_{D-2}^{+++}$ series, which is related  to the heterotic string.

We organise, as in Section~3, the diagonal metric  fields $g_{aa}=
e^{2p^a}\eta_{aa}$ in
two sets:
$p^\alpha$ for  compact dimensions and $p^\mu$ for  non-compact ones in
Minkowski space.  We identify
\begin{eqnarray}
\label{wcompact} p^\alpha&=& \ln
\left(\frac{L_\alpha}{l_{pl}}\right)\qquad \alpha=D,D-1,\dots D-k +1\\
\label{wnoncompact} p^\mu&=&C \qquad\mu=1,\dots  D-k
\end{eqnarray} where $C$ is an arbitrary constant. The embedding of  
${\cal
G}^{(k)}$ in ${\cal G}^{+++}$ is given by Eq.(\ref{compact}).

\subsection{M-theory} Consider the M-theory Lagrangian Eq.(\ref{Mth})
compactified on a
$k$-torus. The connection with the superstrings is obtained by trading  
the
radius $L_{11}$ in  the eleventh dimension and the Planck length
$l_{pl}$ for the string coupling constant $g_s$ and the string length  
$l_s$. One has
\begin{eqnarray}
\label{coupling} L_{11}&=&g_s l_s\, ,\\
\label{length} l_{pl}&=&g_s^{1/3} l_s\, .
\end{eqnarray} The generators of the discrete symmetries $S({\cal
G}^{(k)})\equiv S(E_8^{(k)})\subset S(E_8^{+++})$ are given in this  
case by the
generators of the Weyl reflections only.   The trivial Weyl generators  
of
$E_8^{(k)}$ simply exchange compactification radii  
$L_\alpha\leftrightarrow
L_\beta$. The non-trivial one is the electric Weyl reflection which  
acts  as
\begin{eqnarray}
\label{Weylp2} p^{\prime \alpha}&=&p^\alpha  
+\frac{1}{3}(p^{11}+p^{10}+p^9)
\qquad
\alpha=12-k,\dots 8\quad (k-3 ~{\rm terms}\, ; k\geq 4)\\
\label{Weylp1} p^{\prime \alpha}&=&p^\alpha  
-\frac{2}{3}(p^{11}+p^{10}+p^9)
\qquad
\alpha=9,10,11\\
\label{Weylp3} C^\prime &=& C+ \frac{1}{3}(p^{11}+p^{10}+p^9)\, ,
\end{eqnarray} in accordance with Eqs.(\ref{WeylMe2}) and  
(\ref{WeylMe1}). One
checks that  Eq.(\ref{Weylp3}) also follows for all $k$-tori from
Eqs.(\ref{Weylp2}) and (\ref{Weylp1}),  by using the  equation  
Eq.(\ref{compact})
expressing the embedding of $E_8^{(k)}$ into $E_8^{+++}$.

  From Eqs.(\ref{wcompact}) and (\ref{wnoncompact}) as well as
Eqs.(\ref{coupling}) and (\ref{length}) that relate gravity to string  
parameters,
one gets
\begin{eqnarray}
\label{Tdual} g_s^\prime&=&g_s \frac{l_s^2}{L_{10}L_9}\, ,\nonumber\\
L_{10}^\prime&=&\frac{l_s^2}{L_9}\nonumber\, ,\\
L_9^\prime&=&\frac{l_s^2}{L_{10}}\, ,\nonumber\\
L_\alpha^\prime&=&L_\alpha\qquad \alpha=12-k,\dots 8\qquad (k-3 ~{\rm
terms}\, ; k\geq 4)\, ,
\end{eqnarray} and
\begin{equation}
\label{planck} C^\prime- C= \ln \frac{l_{pl}}{l_{pl}^\prime}\, .
\end{equation} The Eqs.(\ref{Tdual}) describe a double $T$-duality
\cite{dinehs89} for the radii  $L_{10}$ and
$L_9$ together with an exchange of these radii.   We stress that this   
derivation
of the known relation \cite{elitzurgkr97,obersp98,banksfm98} between
$T$-duality and Weyl reflection for M-theory is done by putting on an  
equal
footing the moduli in compact and non-compact dimensions, as needed if  
the
basic symmetry is the very extended group $E_8^{+++}$. The bonus is the
relation Eq.(\ref{planck}) which expresses the rescaling of the  
Minkowskian
metric in the non-compact dimensions when measured in eleven  
dimensions, due
to the change  of the Planck constant under $T$-duality. This  
remarkable result
bares the signature of
$E_{11}=E_8^{+++}$.

\subsection{The bosonic string and the D-series} The relation between  
the string
length and the Planck length in the effective action in $D$ dimensions   
is
\begin{equation}
\label{Dcoupling} l_{pl}=g_s^{\frac{2}{D-2}} l_s\, .
\end{equation} Strictly speaking, this relation only makes sense  for  
$D=26$
where the quantum bosonic string theory is consistent. However,  for  
sake of
generality, we shall leave the dimension $D$ arbitrary and use this  
relation to
define $l_s$.

The moduli $p^i$ in the $L$-basis for the $D$-series are the dilaton
  $\Phi$ and the $p^a=\{ p^\alpha, p^\mu=C\}$. When the theory is  
compactified  on
a
$k$-torus with radii
$L_\alpha$, $p^\alpha$ is equal to $\ln L_\alpha/l_{pl}$ according to
Eq.(\ref{wcompact}) and
$\Phi$ is related to the string coupling constant by
\begin{equation}
\label{couplingD}
\Phi=l \ln g_s\, ,\quad l=\left(\frac{8}{D-2}\right)^{1/2}\, ,
\end{equation} where we have generalised the definition of the  string
coupling\footnote{The coefficient $l$ appearing   in  
Eq.(\ref{couplingD}) is a
consequence of the normalisation for $\Phi$ chosen in the effective  
action
Eq.(\ref{Bth}) as compared to the standard normalisation in string  
theory.}
  to arbitrary
$D$.

As in the M-theory case, we consider $S({\cal G}^{(k)})\equiv
S(D_{D-2}^{(k)})\subset S(D_{D-2}^{+++})$. The trivial Weyl reflections  
in
$S(D_{D-2}^{(k)})$ are associated with the gravity line and simply  
exchange radii.
The electric Weyl reflection follows from   
Eqs.(\ref{Del2})-(\ref{Del0}) with
$a=\{\alpha,\mu\}, ~\alpha=D-k+1,\dots D; \mu=1\dots D-k$.   The  
justification of
using Eq.(\ref{Del2}) when $p^\mu=C$ follows from the embedding equation
Eq.(\ref{compact}) which  yields from Eqs.(\ref {Del1}) and (\ref  
{Del0}).
\begin{equation}
\label{Dweyl} C^\prime=C+{2\over (D-2)}(p^{D-1}+p^D) +{l\over D-2}\Phi
\, ,
\end{equation} in accordance with Eq.(\ref{Del2}) when $a=\mu$.
  Rewriting all equations in terms of
$L_\alpha/l_s$ and $g_s$, and using Eq.(\ref{Dcoupling}), we recover on  
the one
hand,  the double $T$-duality together with the exchange of the first  
two radii
$L_D,L_{D-1}$. On the other hand,  Eq.(\ref{Dweyl}) yields the  
rescaling of the
Minkowskian metric as in Eq.(\ref{planck}).

We now consider the magnetic  Weyl reflection occurring for
$k\ge D-4$.\footnote{Although the magnetic node gets attached to the  
gravity
line at $k=D-3$, it appears at $k=D-4$ (see footnote 5). The magnetic  
Weyl
reflection at $k=D-4$ is defined as the magnetic Weyl reflection  
downlifted from
${\cal G}^{(k)}, k> D-4$.} The moduli transformations are given by
Eqs.(\ref{Dm2})-(\ref{Dm0}) and may be straightforwardly rewritten in  
terms of
the new variables
$p^\alpha$ and
$C$ as before. The equation for $C$ is
\begin{equation}
\label{DMplanck} C^{\prime}=C+{D-4\over D-2}(p^{5}+\dots +p^D) -{l\,  
(D-4)\over
2\, (D-2)}\Phi\, ,
\end{equation} and in terms of the string variables, these read
\begin{eqnarray}
\label{Sdual} g_s^\prime&=&g_s^{-1} \frac{L_D L_{D-1} \dots  
L_5}{l_s^{D-4}}\, ,\\
\frac{L_\alpha^\prime}{l_s^\prime}&=&\frac{L_\alpha}{l_s}
\frac{L_D L_{D-1}  \dots L_5}{l_s^{D-4}} g_s^{-2} \qquad
\alpha=D-k+1,\dots,4~\, (k\geq D-3)\, \label{llb}\\
\frac{L_\alpha^\prime}{l_s^\prime}&=&\frac{L_\alpha}{l_s}
\qquad\qquad\qquad\qquad\qquad
\alpha=5,\dots, D\, .\label{lla}
\end{eqnarray} To ensure that the gravitational constant in four   
dimensions is
invariant under these transformations,
  $l_s$ must transform as
\begin{equation} l_s^\prime = l_s g_s^2 \frac{l_s^{D-4}}{L_D L_{D-1}   
\dots L_5}\, .
\label{lsp}
\end{equation} Recall that in  double $T$-duality no transformation of  
$l_s$ is
needed  to ensure invariance of the gravitational constant.  Inserting
Eq.(\ref{lsp}) in Eqs.(\ref{llb}) and (\ref{lla}) yields
\begin{eqnarray}
  L_\alpha^\prime&=&L_\alpha\qquad\qquad\qquad
\qquad\quad~ \alpha=D-k+1,\dots,4~\, (k\geq D-3)
\label{lld}\\
L_\alpha^\prime &=&L_\alpha \,  g_s^2 \frac{l_s^{D-4}}{L_D
L_{D-1}  \dots L_5}
\qquad
\alpha=5,\dots, D\, .\label{llc}
\end{eqnarray} Note that Eq.(\ref{lld}) implies that the gravitational  
constant is
invariant not only in four,  but also  in three dimensions.  It is  
precisely the
relation Eq.(\ref{lld}) which ensures that the transformation  
Eq.(\ref{DMplanck})
yields the rescaling of the Minkowskian metric  Eq.(\ref{planck}), as  
previously.

We see through Eq.(\ref{Sdual}) that the magnetic Weyl transformation  
reveals in
string language the existence of an $S$-duality  for the bosonic  
string. Such an
$S$-duality has shown up as a symmetry in the coset symmetries of the
dimensional reduction to three dimensions of the heterotic and the    
closed
bosonic string effective action
\cite{julia85}.

Finally, the outer automorphism is read off   
Eqs.(\ref{Dout2})-(\ref{Dout0}) and it
is easily seen that it describes in the string language the simple
$T$-duality on $L_D$, namely $L_D^\prime =l_s^2/L_D~;~ g_s^\prime = g_s
l_s/L_D  $ and yields the required rescaling of Minkowski space.

\subsection{The heterotic string and the B-series}
\vskip-.2cm
As in the D-series, we extend the relation between gravity and string
parameters Eq.(\ref{Dcoupling}) to any space-time dimension $D$ and use  
as
effective Lagrangian the maximally oxidised Lagrangian Eq.(\ref {Hth})  
that
corresponds, when $D=10$, to an heterotic string with one gauge field  
only.

It was pointed out in Section~2.4 that the magnetic Weyl  
transformations of the
$D$- series and the B-series coincide.  Furthermore the Weyl  
transformation
associated to the short root in the Dynkin diagrams of the B-series  
yields the
same result on the physical fields as the outer automorphism of the  
$D$-series.
These transformations describe respectively the $S$-duality and the  
simple
$T$-duality of the heterotic string and relate in a precise manner  the
$S$-duality in the heterotic string and in the bosonic string. As  
previously, the
rescaling of Minkowski space is ensured by the embedding of  
$B_{D-2}^{(k)}$ into
$B_{D-2}^{+++}$.

\section{ Conclusion}
\vskip-.1cm
In this paper we considered the Cartan subalgebra of any very extended  
algebra
${\cal G}^{+++}$ and let the parameters depend on space-time. By   
selecting a
preferred subgroup of ${\cal G}^{+++}$ we identified the space-time  
fields  with
the diagonal metric fields and possibly scalar fields.  One may  view  
this
approach as a special case of the more general procedure  of taking the  
full
non-linear realisation of
${\cal G}^{+++}$ \cite{west01,lambertw01}. Namely, one sets to zero   
all the
fields  of the full non-linear realisation of
${\cal G}^{+++}$ except those associated with the Cartan subalgebra.   
In  the more
general setting, the above identification of fields arises naturally.   
However, in
the full non-linear realisation one has   an infinite number of fields  
corresponding
to the infinite number of generators in ${\cal G}^{+++}$. As there  
exists  no
explicit listing of these latter generators it is not clear how  to  
carry
out this construction in practice. The advantage of the restriction made
in this paper is that one only has a finite number of generators whose  
behaviour
under many    properties of the Kac-Moody algebra  is known and, as a  
result, it is
possible to     perform  calculations completely.

We used this set up to test  gravity  and its coupling in all  oxidised  
theories of
gravity and matter. We found evidence for a symmetry under the  
corresponding
very extended algebra  ${\cal G}^{+++}$, or at least under the group  
$S({\cal
G}^{+++})$ of its Weyl reflections and  outer automorphisms. When  
applied to
M-theory or to  effective actions of superstring or bosonic string  
theories,
these symmetries also appear in the stringy context, thus providing   
indications
that they are  operational at the quantum level as well.  These results  
  provide
evidence, which corroborates the analysis presented in references
\cite{west01,lambertw01,west02,gaberdielw02}, for  the presence of the   
very
extended symmetries
$E_8^{+++}$,
$D_{24}^{+++}$ and
$A_D^{+++}$ in M-theory, the closed bosonic string and pure  gravity
respectively.

We can ask if the appearance of ${\cal G}^{+++}$ symmetries  points  
towards  the
existence of consistent theories of gravity and matter for all ${\cal  
G}$.  It is
useful to recall the classic case of a   non-linear realisation that  
occurs in a
spontaneously broken symmetry in conventional quantum field theory.  
There, at
low energy the dynamics is also given by a non-linear realisation whose  
degrees
of freedom are the Nambu-Goldstone bosons, but the underlying theory
possesses many more degrees of freedom and these  realise linearly the
symmetry.  The prototype example is  the non-linear realisation of pion
dynamics which, as we  know now, is   accounted for by quarks at a more
fundamental level. For those oxidised theories given in  
Eqs.(\ref{Mth}),(\ref{Bth})
and (\ref{Hth}), which can be viewed as effective actions of string  
theories, the
analogy of gravity, dilatons and forms with the Nambu- Goldstone bosons  
of pion
dynamics hiding more fundamental degrees of freedom is borne out by the
existence of the huge number of degrees of freedom in string theory  
itself.
Whether the infinite number of fields in the non-linear realisation  
${\cal G}^{+++}$
could lead to uncovering such new dynamical degrees of freedom and not  
simply
to  a reformulation of the known effective actions, is unclear.     
However,  the
very fact that the realisation would still be a non-linear one  may be  
an
indication that the full theory underlying such  degrees of freedom are  
to be
found in a more elaborate realisation of the ${\cal G}^{+++}$ symmetry.

The analysis of this paper points towards the existence of very extended
symmetries, not only for M-theory and for the bosonic string,  but for  
all
oxidised theories.  The  M-theory quest is generally viewed as the  
privileged
way to reach a unified theory of gravity and matter. From the point of  
view
developed in this paper, the bosonic part of the M-theory effective  
action
Eq.(\ref{Mth}) is just one amongst many sharing the same universal type  
of
symmetry.  Even the simplest oxidised action, namely pure gravity,  
could also be
some kind of effective action hiding many fundamental degrees of  
freedom.
Einstein's theory reveals  the correct thermodynamical entropy of  
Schwarzschild
and Kerr  black holes, but hitherto does not provide an unambiguous  
derivation
of  its statistical content. New degrees of freedom are apparently  
needed to find
the quantum states of the black hole. Although string theory does  
provide such
degrees of freedom and explains successfully the entropy of some black  
holes,
the origin of these degrees of freedom when only gravitational quantum  
numbers
are present might well be  hidden  in the  $A^{+++}$ symmetry.  In  
addition, the
embedding of superstrings into the bosonic string
\cite{casherent85,englertht01} whose effective action is also an  
oxidised theory
Eq.(\ref{Bth}), suggests that the degrees of freedom hidden in  
different $\G$ may
be related to each other.

These considerations motivate the search for structures where the ${\cal
G}^{+++}$ symmetries  could be dynamically realised and which would  
provide
links between different very extended symmetries.

\section*{Acknowledgments}

Fran\c cois Englert and Laurent Houart are very much indebted to Marc  
Henneaux
for explaining to them the structure of cosmological billiards and for  
an
illuminating and constructive discussion on the matter presented in  
Section~4.
Peter West wishes to thank Andrew Pressley for discussions. Laurent  
Houart
would like to thank the CECS  (Centro de Estudios) of Valdivia (Chile)   
and the
Pontificia Universidad Catolica de Chile for the warm hospitality  
extended to him
while this work was finalised. Anne Taormina and Peter West thank the
Universit\'e Libre de Bruxelles for the friendly and stimulating  
atmosphere
during their visits.

This work was supported in part  by the NATO grant PST.CLG.979008,
   by the ``Actions de Recherche Concert\'ees'' of the ``Direction de la  
Recherche
Scientifique - Communaut\'e Fran\c caise de Belgique, by a ``P\^ole  
d'Attraction
Interuniversitaire'' (Belgium), by IISN-Belgium (convention 4.4505.86),  
by
Proyectos FONDECYT 1020629, 1020832 and 7020832 (Chile) and by the  
European
Commission RTN programme HPRN-CT00131, in which F.~E and L.~H. are
associated to the Katholieke Universiteit te Leuven (Belgium).


\begin{thebibliography}{99}
\bibitem{campbellw84} C. Campbell and P. West, {\it $N=2$ $D=10$  
non-chiral
supergravity and its spontaneous compactification}, Nucl.\ Phys.\ {\bf  
B243}
(1984) 112.

\bibitem{huqn85} M. Huq and M. Namazie, {\it Kaluza--Klein supergravity  
in ten
dimensions}, Class.\ Quant.\ Grav.\ {\bf 2} (1985) 293.

\bibitem{gianip84} F. Giani and M. Pernici, {\it $N=2$ supergravity in  
ten
dimensions}, Phys.\ Rev.\ {\bf D30} (1984) 325.

\bibitem{schwarzw83} J. Schwarz and P. West, {\it Symmetries and
transformations of chiral
$N=2$ $D=10$ Supergravity}, Phys. Lett. {\bf 126B} (1983) 301.

\bibitem {howew84} P. Howe and P. West, {\it The Complete $N=2$ $D=10$
supergravity}, Nucl.\ Phys.\ {\bf B238} (1984) 181.

\bibitem {schwarz83} J. Schwarz, {\it Covariant field equations of  
chiral $N=2$
$D=10$ supergravity}, Nucl.\ Phys.\ {\bf B226} (1983) 269.

\bibitem{brinkss77} L. Brink, J. Scherk and J.H. Schwarz, {\it  
Supersymmetric
Yang-Mills theories}, Nucl. Phys. {\bf B121} (1977) 77; F. Gliozzi, J.  
Scherk and D.
Olive, {\it Supersymmetry, supergravity theories and the dual spinor  
model},
Nucl. Phys. {\bf B122} (1977) 253;  A.H. Chamseddine, {\it Interacting  
supergravity
in ten dimensions: the role of the six-index gauge field}, Phys. Rev.  
{\bf D24}
(1981) 3065; E.\ Bergshoeff, M.\ de Roo, B.\ de Wit and P.\ van  
Nieuwenhuizen, {\it
Ten-dimensional Maxwell-Einstein supergravity, its currents, and the  
issue of its
auxiliary fields}, Nucl.\ Phys.\ {\bf B195} (1982) 97; E.\ Bergshoeff,  
M.\ de Roo and
B.\ de Wit, {\it Conformal supergravity in ten dimensions}, Nucl.\  
Phys.\ {\bf B217}
(1983) 143;  G. Chapline and N.S. Manton, {\it Unification of  
Yang-Mills theory and
supergravity in ten dimensions}, Phys. Lett. {\bf 120B} (1983) 105.

\bibitem{cremmerjs78} E. Cremmer, B. Julia and J. Scherk,  {\it  
Supergravity
theory in eleven dimensions}, Phys. Lett. {\bf 76B} (1978) 409.

\bibitem{townsend95}  P.K. Townsend, {\it The eleven dimensional
supermembrane revisited}, Phys.\ Lett.\ {\bf B350} (1995) 184, {\tt
arXiv:hep-th/9501068}; {\it D-branes from  M-branes}, Phys.\ Lett.\  
{\bf B373}
(1996) 68, {\tt arXiv:hep-th/9512062}.

\bibitem{witten95} E. Witten, {\it String theory dynamics in various  
dimensions},
Nucl.\ Phys.\ {\bf B443} (1995) 85, {\tt arXiv:hep-th/9503124}.

\bibitem {west00} P. West, {\it Hidden superconformal symmetries of  
M-theory},
{\bf JHEP 0008} (2000) 007, {\tt arXiv:hep-th/0005270}.

\bibitem{west01} P. West, {\it $E_{11}$ and M Theory}, Class. Quant.  
Grav.  {\bf 18}
(2001) 4443, {\tt arXiv:hep-th/ 0104081}.

\bibitem{lambertw01} N. Lambert and  P. West, {\it Coset symmetries in
dimensionally reduced bosonic string theory}, Nucl. Phys. {\bf B615}  
(2001) 117,
{\tt arXiv:hep-th/0107209}.

\bibitem{goddardo84} P. Goddard and D. Olive, {\it Algebras, lattices  
and strings,}
in {\it Vertex operators in Mathematics and Physics}, MSRI Publication  
no 3,
Springer (1984) 51.

\bibitem{olivegw02} D. Olive, M. Gaberdiel and P. West,  {\it A class  
of Lorentzian
Kac-Moody algebras}, Nucl. Phys. {\bf B645} (2002) 403, {\tt
arXiv:hep-th/0205068}.



\bibitem{damourhn00} T. Damour, M. Henneaux and H. Nicolai, {\it  
Cosmological
billiards}, {\tt arXiv:hep-th/ 0212256}.

\bibitem{cremmerjlp99} E. Cremmer, B. Julia, H. Lu and C. Pope, {\it  
Higher
dimensional origin of $D=3$ coset symmetries}, {\tt  
arXiv:hep-th/9909099}.

\bibitem{ms} N. Marcus and J. Schwarz, {\it Three-dimensional
supergravity theories}, Nucl. Phys. {\bf B228} (1983) 301. 


\bibitem{damourbhs02} T. Damour, S. de Buyl, M. Henneaux and C.  
Schomblond,  {\it
Einstein billiards and overextensions of finite-dimensional simple Lie  
algebras},
{\bf JHEP 0208} (2002) 030, {\tt arXiv:hep-th/0206125}.


\bibitem{ferrarasz77} S.\ Ferrara, J.\ Scherk and B.\ Zumino, {\it  
Algebraic
properties of extended supersymmetry}, Nucl.\ Phys.\ {\bf B121} (1977)  
393; E.\
Cremmer, J.\ Scherk and S.\ Ferrara, {\it SU(4) Invariant supergravity  
theory},
Phys.\ Lett.\ {\bf 74B} (1978) 61.

\bibitem{cremmerj78} E. Cremmer and B. Julia, {\it The $N=8$  
supergravity
theory. I. The Lagrangian.}, Phys.\ Lett.\ {\bf 80B} (1978) 48.

\bibitem{julia81} B.\ Julia, {\it Group disintegrations}, in {\it  
Superspace and
Supergravity}, p. 331,  eds. S. W. Hawking  and M.  Ro\v{c}ek,  
Cambridge University
Press (1981).

\bibitem{julia98} B. Julia, {\it Dualities in the classical supergravity
limits}, {\tt   arXiv:hep-th/9805083}. 

\bibitem{nicolai87}  H. Nicolai, {\it The integrability of N=16  
supergravity},
  Phys. Lett. {\bf 194B} (1987) 402; {\it On M-Theory}, {\tt  
arXiv:hep-th/9801090}.

\bibitem {julia84} B. Julia, in {\it Vertex Operators in Mathematics  
and Physics},
Publications of the Mathematical Sciences Research Institute no 3,  
Springer
Verlag (1984); B. Julia in {\it Superspace and Supergravity} edited by  
S.W.
Hawking and M. Rocek, Cambridge University Press (1981).






\bibitem{kikkaway84}  K. Kikkawa and M. Yamasaki, {\it Casimir effects  
in
superstring theories}, Phys. Lett. {\bf B149} (1984) 357; N. Sakai and  
I. Senda, {\it
Vacuum energies of string compactified on torus}, Prog. Theor. Phys.  
{\bf 75}
(1986) 692; T. H. Buscher, {\it A symmetry of the string background  
field
equations}, Phys. Lett. {\bf B194} (1987) 59; {\it Path integral  
derivation of
quantum duality in non-linear $\sigma$-models}, Phys. Lett. {\bf B201}  
(1988) 466.

\bibitem{rocekv92}  M. Rocek and E. Verlinde, {\it Duality, quotients  
and
currents}, Nucl. Phys. {\bf B373} (1992) 630, {\tt  
arXiv:hep-th/9110053}.

\bibitem{fontilq90} A. Font, L. Ibanez, D. Lust and F. Quevedo, {\it  
Strong-weak
coupling duality and nonperturbative effects in string theory},   
Phys.Lett. B249
(1990) 35; S. J. Rey, {\it The confining phase of superstrings and  
axionic strings},
Phys. Rev. {\bf D43} (1991) 526.

\bibitem{hullt94} C.M. Hull and P.K. Townsend, {\it Unity of  
superstring  dualities},
Nucl.\ Phys.\ {\bf B438} (1995) 109, {\tt arXiv:hep-th/9410167}.

\bibitem{elitzurgkr97} S. Elitzur, A. Giveon, D. Kutasov and E.  
Rabinovici,  {\it
Algebraic aspects of matrix theory on $T^d$ }, {\tt  
arXiv:hep-th/9707217}.

\bibitem{obersp98} N. Obers and B. Pioline,~ {\it U-duality and  
M-theory, an
algebraic approach}~, {\tt arXiv:hep-th/9812139}.

\bibitem{banksfm98} T. Banks and W. Fischler and L. Motl, {\it  
Dualities versus
singularities}, {\bf JHEP 9901} (1999) 019, {\tt arXiv:hep-th/9811194}.

\bibitem{moore93} G. Moore, {\it Finite in all directions}, {\tt
arXiv:hep-th/9305139};
  P. West, {\it Physical states and string symmetries}, Mod. Phys. Lett.  
{\bf A10}
(1995) 761, {\tt arXiv:hep-th/9411029}.

\bibitem{harveym95} J. Harvey and G. Moore, {\it On the algebra of BPS  
states},
Nucl. Phys. {\bf B463} (1996) 315, {\tt arXiv:hep-th/9510182}; {\it  
Exact
gravitational threshold corrections in the FHSV model}, Phys. Rev. {\bf  
D57}
(1998) 2329, {\tt arXiv:hep-th/9611176}.

\bibitem {dewitn86} B. de Wit and H. Nicolai, {\it D=11 supergravity  
with local
$SU(8)$ invariance}, Nucl. Phys. {\bf B274} (1986) 363; H.  Nicolai,  
{\it Hidden
symmetries in D=11 supergravity},  Phys. Lett. {\bf 155B} (1985) 47;   
H.  Nicolai,
{\it D=11 supergravity with local $SO(16)$ invariance},  Phys. Lett.  
{\bf 187B}
(1987) 316.

\bibitem{meloshn97}   S. Melosch and H. Nicolai, {\it New canonical  
variables for
$D=11$ supergravity}, Phys. Lett. {\bf B416} (1998) 91, {\tt
arXiv:hep-th/9709227};
  K. Koepsell, H. Nicolai and H. Samtleben, {\it An Exceptional   
Geometry for $d=11$
Supergravity}, Class. Quant. Grav. {\bf 17} (2000) 3689, {\tt
arXiv:hep-th/0006034}; B. de Wit and H. Nicolai, {\it Hidden  
symmetries, central
charges and all that},  Class. Quant. Grav. {\bf 18} (2001)  3095, {\tt
arXiv:hep-th/0011239}.

\bibitem{cremmerjlp98} E. Cremmer, B. Julia, H. Lu and C. Pope, {\it  
Dualisation of
dualities II: Twisted self-duality of doubled fields and  
superdualities}, Nucl. Phys.
{\bf B535} (1988) 242, {\tt arXiv: hep-th/9806106}.

\bibitem{damourh00}  T. Damour, M. Henneaux, {\it Chaos in superstring
cosmology}, Phys. Rev. Lett. 85 (2000) 920, {\tt arXiv:hep-th/0003139};
  T. Damour, M. Henneaux, {\it E(10), BE(10) and arithmetical chaos in  
superstring
cosmology}, Phys. Rev.Lett. {\bf 86} (2001) 4749, {\tt  
arXiv:hep-th/0012172}.

\bibitem{damourhjn01} T. Damour, M. Henneaux, B. Julia and H. Nicolai,  
{\it
Hyperbolic Kac-Moody algebras and chaos in Kaluza-Klein models}, Phys.  
Lett.
{\bf B509} (2001) 323,  {\tt arXiv: hep-th/0103094}.







\bibitem{schnakw01} I. Schnakenburg and  P. West, {\it Kac-Moody   
symmetries of
IIB supergravity}, Phys. Lett. {\bf B517} (2001) 421, {\tt  
arXiv:hep-th/0107181}.

\bibitem{damourhn02} T. Damour, M. Henneaux and H. Nicolai, {\it  
$E_{10}$ and a
small tension expansion of M-theory}, Phys. Rev. Lett. {\bf 89} (2002)  
221601, {\tt
arXiv:hep-th/0207267}.

\bibitem{henryjp02} P. Henry-Labordere, B. Julia and L. Paulot, {\it  
Borcherds
symmetries in M-theory}, {\bf JHEP 0204} (2002) 049, {\tt  
arXiv:hep-th/0203070}.

\bibitem{casherent85} A. Casher, F. Englert, H. Nicolai and A.  
Taormina, {\it
Consistent superstrings as solutions of the D=26 bosonic string theory}  
Phys.
Lett {\bf B162 } (1985) 121; F. Englert, H. Nicolai, A. Schellekens,   
{\it
Superstrings from $26$ dimensions},   Nucl. Phys. {\bf B274} (1986) 315.

\bibitem{englertht01} F. Englert, L. Houart and  A. Taormina, {\it    
Brane fusion in
the bosonic string and the emergence of fermionic strings},  {\bf JHEP  
0108}
(2001) 013, {\tt arXiv:hep-th/0106235};  A. Chattaraputi,  F. Englert,  
L. Houart and
A. Taormina, {\it The bosonic mother of fermionic D-branes }, {\bf JHEP  
0209}
(2002) 037,  {\tt arXiv:hep-th/0207238}.

\bibitem{borisovo74} A. Borisov and V. Ogievetsky,  {\it Theory of  
dynamical
affine and conformal  symmetries as the theory of the gravitational  
field},  Teor.
Mat. Fiz. {\bf 21} (1974) 329.

\bibitem {kac83}  V. Kac, {\it Infinite dimensional Lie algebras},  
Birkhauser, 1983.

\bibitem{fradkint85}  E. Fradkin and A. Tseytlin, {\it Quantum string  
theory
effective action}, Nucl Phys. {\bf B261} (1985) 1; {\it Effective field  
theory from
quantized strings}, Phys. Lett. {\bf B158} (1985) 316.

\bibitem{west02} P. West, {\it Very Extended $E_8$ and $A_8$ at low  
levels,
gravity and supergravity}, {\tt arXiv: hep-th/0212291}.

\bibitem{booklifschitz} L. Landau and E. Lifschitz, {\it  Th\'eorie du  
champ},
(Traduit du russe par E. Gloukhian), Edition de la paix, Moscou,  
Chapter 11, p. 410.

  \bibitem{giveonpr}A. Giveon, M. Porrati and E. Rabinovici,  {\it  
Target space
duality in string theory},  Physics Reports {\bf 244} (1994) 77, {\tt
arXiv:hep-th/9401139}.

\bibitem{dinehs89} M. Dine, P. Huet and N. Seiberg, {\it Large and  
small radius in
string theory}, Nucl.\ Phys.\ {\bf B322} (1989) 301;  J. Dai, R.G.  
Leigh, J. Polchinski,
{\it New connections between string theories}, Mod.\ Phys.\ Lett.\ {\bf  
A4} (1989)
2073;  E. Bergshoeff, C. Hull and T. Ortin, {\it  Duality in the  
type-II superstring
effective action}, Nucl.\ Phys.\ {\bf B451} (1995) 547, {\tt  
arXiv:hep-th/9504081}.

\bibitem{julia85} B. Julia, {\it Kac-Moody symmetry of gravitation and
supergravity theories}, Lectures in Applied Mathematics, Vol. {\bf 21}  
(1985) 335;
A. Sen, {\it Strong-weak coupling duality in four dimensional string  
theory},  Int.
J. Mod. Phys. {\bf A9} (1994) 3707, {\tt arXiv:hep-th/9402002}; {\it  
Strong-weak
coupling duality in three dimensional string theory}, Nucl. Phys. {\bf  
B434} (1995)
179, {\tt arXiv:hep-th/9408083}.

\bibitem{gaberdielw02} M. Gaberdiel and P. West, {\it Kac-Moody  
algebras in
perturbative string theory}, {\bf JHEP 0208} (2002) 049, {\tt
arXiv:hep-th/0207032}.



\end{thebibliography}
\end{document}